\documentstyle[epsfig,12pt]{article}
\newcommand{\degree}{\hbox{$^\circ$}}
\begin{document}

\begin{center}
\Large {\bf The Design and Performance of the ZEUS Central
Tracking Detector $z$-by-Timing System}
\end{center}

\begin{flushleft}
\vspace{10mm}
D.S.~Bailey, B.~Foster, G.P.~Heath, 
C.J.S.~Morgado\footnote{Now at M.A.I.D. PLC., 38 Leicester Square, 
London WC2H~7DB, UK.}. \\
{\footnotesize\it University of Bristol, H.H.~Wills Physics Laboratory, 
Tyndall Avenue, Bristol BS8~1TL, UK.} \\

\bigskip
N.~Harnew, T.~Khatri\footnote{Now at A.T.~Kearney Ltd., 130 Wilton Road,
London SW1V~1LQ, UK.}, 
M.~Lancaster\footnote{Now at LBL, Berkeley, CA94720, USA.}, 
I.C.~McArthur, J.D.~McFall,
J.~Nash\footnote{Now at Unipart Information Technology, Cowley, 
Oxford OX4~2PG, UK.}, P.D.~Shield, S.~Topp-Jorgensen, 
F.F.~Wilson\footnote{Now at PPE Division, 
CERN, 1211 Geneva 23, Switzerland.}.  \\
{\footnotesize\it University of Oxford, Department of Nuclear Physics, Keble 
Road, Oxford OX1~3RH, UK.} \\

\bigskip
R.C.~Carter, M.D.~Jeffs, R.~Milborrow, M.C.~Morrissey, 
D.A.~Phillips\footnote{Now at Data \& Research Services PLC., 
Linford Road, Milton Keynes MK14~6LR, UK.}, 
S.P.H.~Quinton, G.~Westlake, D.J.~White.  \\
{\footnotesize\it Rutherford Appleton Laboratory, Chilton, Didcot, Oxon 
OX11~0QX, UK.} \\

\bigskip
J.B.~Lane, G.~Nixon, M.~Postranecky. \\
{\footnotesize\it University College, Department of Physics and Astronomy,
Gower Street, London WC1E~6BT, UK.} \\

\end{flushleft}

\vspace{5mm}
\begin{abstract} 

The ZEUS Central Tracking Detector utilizes a time difference measurement
to provide a fast determination of the $z$~coordinate of each hit.
The $z$-by-timing measurement is achieved by using
a Time-to-Amplitude Converter which has an intrinsic timing resolution
of 36~ps, has pipelined readout, and has a multihit
capability of 48~ns. In order to maintain 
the required sub-nanosecond timing accuracy,
the technique incorporates an automated self-calibration system.
The readout of the $z$-by-timing data utilizes a fully customized timing 
control system which runs synchronously with the HERA beam-crossing 
clock, and a data acquisition system implemented
on a network of Transputers.
Three dimensional space-points provided by the $z$-by-timing system
are used as input to all three levels of the ZEUS trigger and for
offline track reconstruction.   The average $z$~resolution
is determined to be
4.4~cm for multi-track events from positron-proton collisions in the
ZEUS detector.

\end{abstract}
\setcounter{footnote}{0}

\newpage
\section{Introduction}
\label{sec:Intro}

This paper describes the readout and performance of the ZEUS Central 
Tracking Detector $z$-by-timing system. ZEUS is a multipurpose
detector~\cite{ZEUS} which records collisions
of 27.5~GeV electrons or positrons 
with 820~GeV protons at the Hadron Electron Ring Accelerator (HERA) at DESY. 
The Central Tracking Detector (CTD)~\cite{CTD1,CTD2} is a
cylindrical drift chamber which surrounds the beam
pipe at the $e-p$ interaction region and is located directly inside a
superconducting magnet which provides a 1.43~T axial field. 
Its purpose is to measure momenta of charged particles 
between polar angles of 7.5$\degree<\theta <164\degree$, provide
$dE/dx$ information to enhance particle identification and to provide
tracking information to all three levels of the ZEUS trigger.

The objective of
the ZEUS First Level Trigger~(FLT)~\cite{ZEUS,CTD_trigger}  is to 
enhance the sample of genuine electron-proton collisions by reducing the
30~kHz background rate from proton beam-gas 
and beam-scraping interactions down to a maximum of 1~kHz.
The CTD Track Trigger is a component of
the ZEUS FLT which contributes to this reduction. Firstly it identifies 
those events which have a vertex consistent with 
originating from the $e-p$ interaction region. Secondly 
it can select interesting physics processes that may 
deposit only small amounts of energy in the ZEUS calorimeter but 
have a distinctive track topology
(e.g. elastic $\rho^0$ and $J/\psi$ production~\cite{rhopaper,jpaper}).

The CTD has a multi-cell stereo superlayer wire geometry, similar in design to 
the CDF chamber~\cite{CDF}. 
Five of nine superlayers are axial, having 
wires running parallel to the $z$~axis\footnote{
The ZEUS coordinate system is a right handed system with the $z$~axis
pointing in the proton beam direction, the $x$~axis pointing 
to the centre of HERA and the $y$ axis vertically upwards.  Hereafter we
refer to the proton ($+z$) direction as the `forward' direction
and the electron ($-z$) direction as the `rear' direction.};
the remaining four superlayers have $\sim$5$\degree$ stereo 
angles which allow three-dimensional track reconstruction
of the $z$~coordinate. 
The stereo information from the CTD
provides an accurate offline measurement,
however it cannot be used in the ZEUS FLT because
of insufficient processing time.
Hence for the FLT we have chosen the technique of $z$-by-timing which can
provide fast (in our case a digitization within 48~ns), 
three-dimensional space-point measurements of the CTD hit information.
The method utilizes a time difference measurement in which the $z$~coordinate 
is proportional to the time difference between the induced pulse arriving at
each end of the wire,
the constant of proportionality being nominally half the speed of
light~\cite{zbyt_1,zbyt_2,zbyt_3}. 
By comparison, measuring the $z$~position by the method of charge
division would be too slow because of the need
to integrate charge over many tens of nanoseconds.

HERA is designed to run with 220 bunches in each of the electron and
proton rings with a time interval between crossings of 96~ns.
The 96~ns beam crossing interval 
puts special demands on triggering and readout of the ZEUS detector.
To ensure that ZEUS is sensitive to every  HERA 
beam crossing, the data through the $z$-by-timing system
and the FLT processors must be pipelined~\cite{ZEUS,zbyt_1}.
This requirement is the overriding influence on the design
of the readout architecture and trigger system.

As well as being an important component of the ZEUS FLT,
the $z$-by-timing information is also used as input to 
trigger levels 2 and 3. The Second Level 
Trigger~(SLT)~\cite{ZEUS,CTD_SLT} reduces the maximum 1~kHz First Level
rate down to a maximum of 100~Hz by refined processing
of track and calorimeter information.
Here  the $z$-by-timing information is used as input to a track finding and
vertex fitting package, implemented on a network of Transputers. 
At the Third Level Trigger (TLT) 
the space-points provided by the $z$-by-timing
system are used in full track reconstruction to obtain an online vertex
measurement (giving a stand-alone resolution of $\sim$40~mm). 
The output from the TLT is written to magnetic storage at 
a rate of $\sim$3-5~Hz. 

Offline, not only does the 
$z$-by-timing system provide an aid to track finding at the first stage  
of track reconstruction but it can, and has been,
used as an independent tracking system in its own right.
During the first year of HERA data-taking, the 
$z$-by-timing system provided the sole tracking readout system in ZEUS. 
Throughout the period of HERA operation,
the CTD and the $z$-by-timing system have run successfully and
reliably.

This paper is outlined as follows.
The CTD is described in Section~\ref{sec:CTD}
and an overview of its readout is given in Section~\ref{sec:CTD_RO}. 
Section~\ref{sec:amps} describes the CTD amplification system
and the implications for the linearity of the
$z$-by-timing measurement.
The $z$-by-timing readout cards, which  incorporate
time-to-amplitude conversion and digitization, are 
described in Section~\ref{sec:ZbyT}. 
Section~\ref{sec:cal} discusses the CTD calibration system, including 
pulse generation and calibration control. Section~\ref{sec:DAQ} describes 
timing control and data acquisition (DAQ) of the $z$-by-timing data.
The performance during the first three years
of HERA operation is detailed in Section~\ref{sec:results}. 
Finally, Section~\ref{sec:summary} contains
a brief summary.

\section{The Central Tracking Detector}
\label{sec:CTD}
\setcounter{footnote}{0}

The CTD is one of the inner tracking detectors of ZEUS. It is a
cylindrical drift chamber with inner and outer active radii of 190~mm and
785~mm respectively, and has an active length of 2.03~m. It surrounds the 
beam pipe at the $e-p$ interaction region and is located inside a
superconducting solenoid which provides a 1.43~T 
axial magnetic field. The chamber axis is coincident with the beam axis.

\begin{figure}[tb]
\begin{center}
\epsfig{file=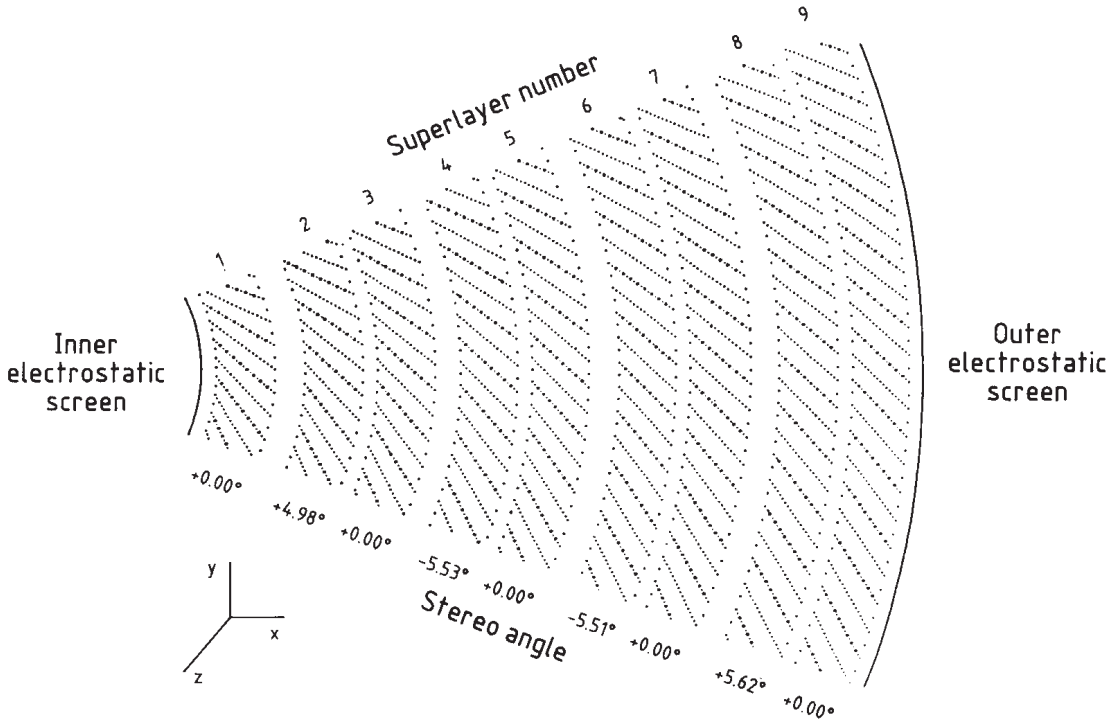,height=10cm}
\caption{ The layout of an octant of the CTD at the chamber
endplate. Sense wires are indicated by highlighted dots.}
\vfill
\label{fig:ctdend} 
\end{center}
\end{figure}

Fig.~\ref{fig:ctdend} shows an octant of the CTD endplate normal to the beam
axis in the $r-\phi$ projection\footnote{Here $r$ 
is defined as the radial
distance from the $z$~axis, $\phi$ is the azimuthal angle measured
with respect to the $x$ axis. The polar angle $\theta$ is defined
with respect to the proton beam direction.}.
The total number of wires in the detector is 24192
of which 4608 are sense wires. The wires are arranged in nine `superlayers'.
The five odd-numbered superlayers are axial, their wires running parallel to
the beam axis. The four even-numbered superlayers
have small stereo angles, their
wires being skewed in $\phi$ by approximately $\pm$5$\degree$. This
combination of axial and stereo 
superlayers allows accurate three-dimensional
reconstruction of charged particle tracks at the offline analysis stage. 

A superlayer is divided into a number of `cells', each containing eight
sense wires. The cell boundary is defined by planes of field wires. The eight
sense wires, separated by field-shaping wires held at ground potential, are
arranged in a straight line at 45$\degree$ to a radial line from the chamber
axis. This arrangement has a number of advantages for the chamber
performance~\cite{CTD1,CDF}. 
The separation of adjacent sense wires is approximately
8~mm and the distance from the sense wires to the cell boundary, which defines
the maximum drift distance in the cell, is typically 25~mm. The sense wires
are tungsten and have a diameter of 30~$\mu$m and a resistance of
80~$\Omega$/m. 

A total of 704 wires are instrumented with $z$-by-timing readout.
All wires in Superlayer~1 are equipped in order to give maximum
acceptance for tracks in the forward proton direction
(note that tracks from high $Q^2$ $e-p$ collisions are 
predominantly produced in the
forward proton direction due to the asymmetric beam energies, whilst
the background is similarly forward since it arises from proton
collisions with stationary gas molecules). For less
forward-going tracks, the lever arm, and hence the vertex determination, is 
greatly improved by adding a relatively small 
number of hits from additional axial superlayers.
Hence as a compromise between improved performance and cost,
axial Superlayers~3 and 5 are half instrumented with readout.

The voltages on the wires in a cell are used to define an operational
drift field ($E_d$) and sense wire surface field ($E_s$). 
The choice of these parameters
is dependent on the amplifying gas mix in the chamber
which in turn is an important consideration for efficient and
safe operation of the CTD. 
We have operated the CTD and its prototypes
with a variety of gas mixes at 3~mbar above atmospheric pressure, and
their constitutions are given in Table~\ref{table:gases}.
All gases and operating conditions listed 
result in azimuthal drift with velocity $\sim$50~$\mu$m/ns,
a maximum drift time of 500~ns (equivalent to
five beam crossings of the HERA accelerator) and
a sense wire surface field which defines
a gas gain of approximately 1$\times$10$^5$~\cite{Nash_Salmon}.
To ensure the condition of
azimuthal drift, the tilted cell geometry requires the choice
of a 45$\degree$ Lorentz angle.
Our preferred gas mix is 50:50 argon:ethane with a small admixture
of ethanol ($\sim$1.6\%).  
With a 1.6~kV/cm drift field this gives a 45$\degree$ Lorentz angle in a
1.8~T magnetic field, the nominal field of ZEUS.
Unfortunately technical difficulties have limited the field to 1.43~T
which has put stringent constraints on  chamber operation.
In addition, to date we have
chosen to operate the CTD with  `safe' argon:CO$_2$:ethane gas mixes
during the relatively low luminosity running of HERA. 
This has meant we have had to run the CTD at a drift field close to
1.2~kV/cm, below which the field shaping becomes poor~\cite{CTD1}. This has
implications on chamber performance, especially on the hit efficiency
at cell boundaries. We will return to this point in 
Section~\ref{sec:results}. 

\begin{table}[htb]
\begin{center}
\small{
\begin{tabular}{|c|c|c|c|c|c|c|c|c|c|}
\hline
\multicolumn{3} {|c|}{ } &  &  &  &  &  & &  \\
\multicolumn{3} {|c|}{Nominal gas } & Added &  &  &  &  &  & Year of \\
\multicolumn{3} {|c|}{mix by volume (\%) } & Ethanol & & $E_d$ & 
$E_s$ & $v_d$ & $\theta_{Lorentz}$ & HERA \\
\cline{1-3}
 &  &  & Fraction & $B$ (T) & (kV/cm) & (kV/cm) & 
($\mu$m/ns) & (degrees) & operation \\
Argon & CO$_2$ & Ethane & (\%) &  &  &  & ($\pm$1\%) & ($\pm$0.5\%) & \\
  &  &  &  &  &  &  &  &  & \\
\hline
  &  &  &  &  &  &  &  &  & \\
50 & 0 & 50 & 1.6 & 1.80 & 1.50 & 232 & 51.0 & 45.0 & T.B. \\
  &  &  &  &  &  &  &  &  & \\
85 & 13 & 2 & 0.95 & 1.80 & 1.35 & 192 & 49.0 & 45.0 & T.B. \\
  &  &  &  &  &  &  &  &  & \\
90 & 8 & 2 & 0.84 & 1.43 & 1.20 & 172 & 47.3 & 44.7 & 1992/3 \\
  &  &  &  &  &  &  &  &  & \\
85 & 8 & 7 & 0.84 & 1.43 & 1.22 & 182/4 & 50.5 & 43.4 & 1994 \\
  &  &  &  &  &  &  &  &  & \\
83 & 5 & 12 & 0.84 & 1.43 & 1.22 & 183 & 48.3 & 44.7 & 1995/6 \\
  &  &  &  &  &  &  &  &  & \\
\hline
\end{tabular}
}
\end{center}
\caption{\label{table:gases}  The nominal operating conditions for
various CTD gas mixes which have been used over the course of the
HERA running period
(T.B. signifies test beam operation prior to HERA startup). 
$E_d$ and $E_s$ are the drift field and 
sense wire surface fields respectively which result in a drift velocity
$v_d$ and Lorentz angle $\theta_{Lorentz}$ in
an applied magnetic field $B$.}
\vfill
\end{table}

\section{Overview of the CTD Readout}
\label{sec:CTD_RO}

\begin{figure}[htb]
\begin{center}
\epsfig{file=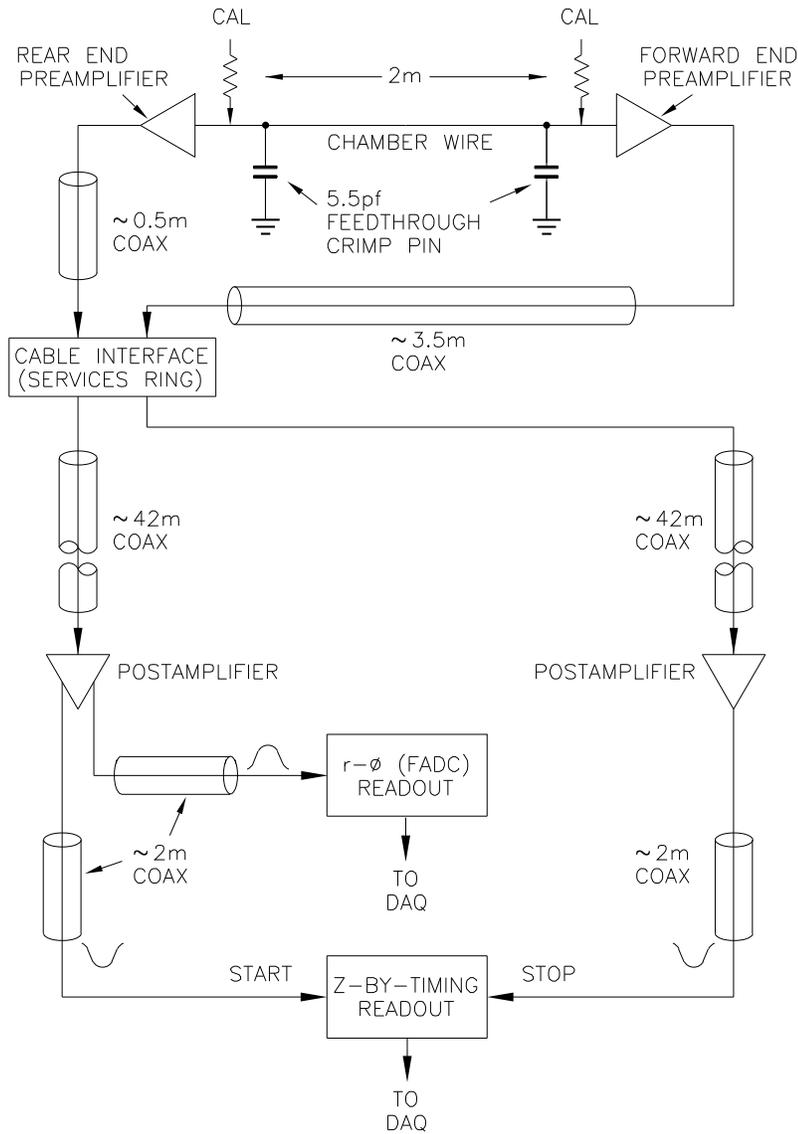,height=15cm}
\caption{ A block diagram of the CTD front-end readout system.}
\vfill
\label{fig:block} 
\end{center}
\end{figure}

Fig.~\ref{fig:block} shows a block diagram of the CTD front-end readout
system.
The major components are as follows:

\begin{itemize}

\item{ } Preamplifiers mounted directly on either
end of the CTD end-flanges amplify the chamber pulses
and drive the signals down 42~m of coaxial cable to postamplifiers,
which provide most of the electronic gain.

\smallskip
\item{ } A Flash-Analogue-to-Digital Converter (FADC) system clocked
at 104~MHz  (hereafter referred to as the $r-\phi$ FADC system)
samples the amplified pulses from
all 4608 sense wires and measures drift times~\cite{FADC,Cussans_etal}.
In this way, $r-\phi$ trajectories of charged
particles in the plane perpendicular to the beam
are measured to a precision of about 190~$\mu$m on each wire.
The performance of this readout system will be the subject of another
paper.

\smallskip
\item{ } The $z$-by-timing readout system 
digitizes the time difference between the arrival of the 
pulses from the two ends of the chamber using a
pipelined Time-to-Amplitude Converter (TAC). The intrinsic
resolution of the TAC is $\sim$36~ps. 

\end{itemize}

\begin{figure}[htb]
\begin{center}
\epsfig{file=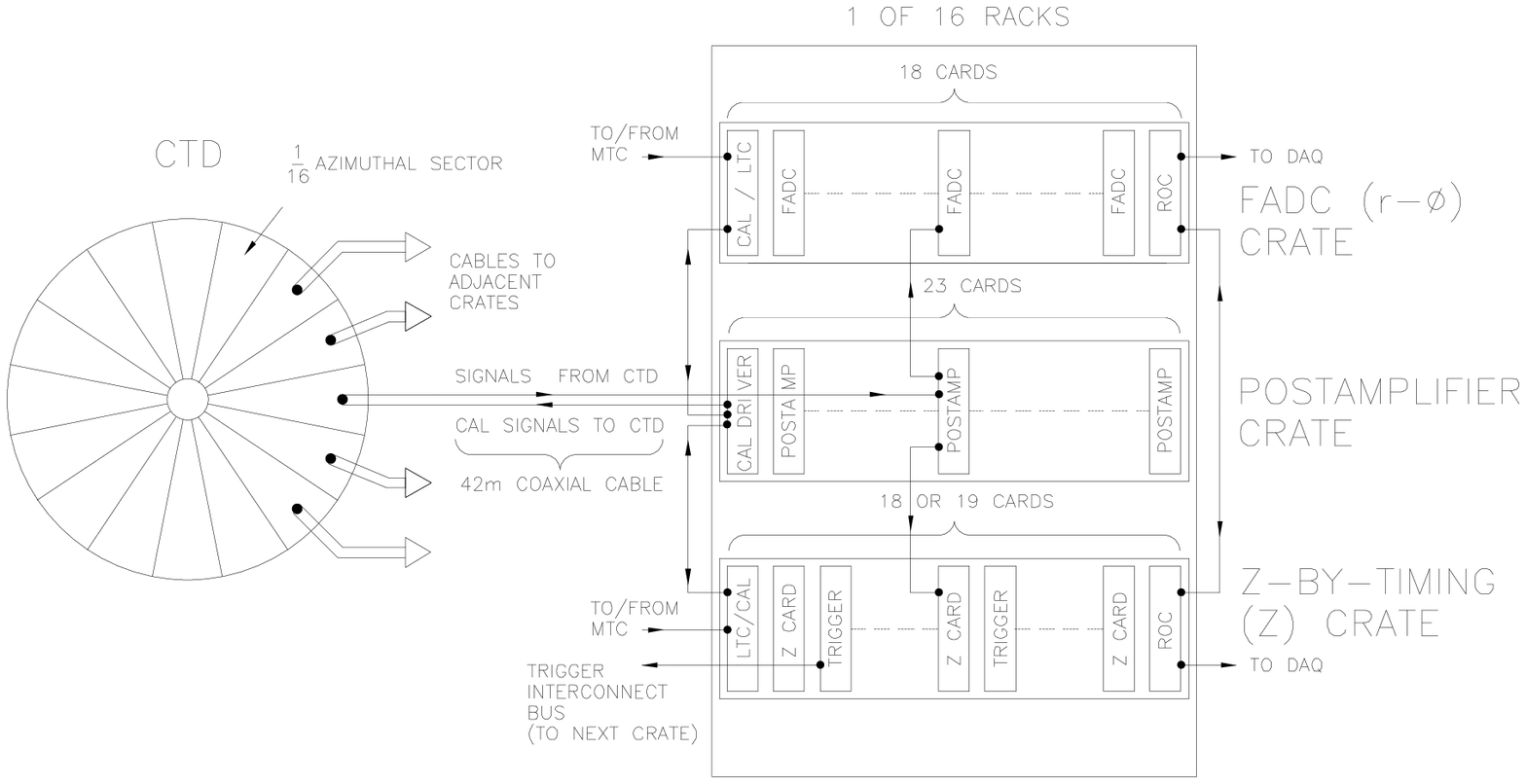,height=8cm}
\caption{ The crate layout for ${{1}\over{16}}^{th}$ of the CTD.}
\vfill
\label{fig:crate_layout}
\end{center}
\end{figure}

The readout philosophy of the CTD utilizes the geometrical symmetry
of the chamber. We have chosen to electronically divide the CTD into 
16 azimuthal sectors in $\phi$, each of which is serviced independently
by its own rack of electronics. 
The electronics for each sector
are contained in three crates, stacked within a single rack.
Fig.~\ref{fig:crate_layout} demonstrates the layout.
The `Postamplifier Crate' contains postamplifier cards for that
sector. From the postamplifiers the 
pulses are split into the 104~MHz FADCs
located in the `$r-\phi$~Crate', and the $z$-by-timing
electronic cards, located in the `$z$-Crate'. 
These are custom-built 9-Unit (9U) high $Teradyne$ crates~\cite{Teradyne}.

Each $z$-crate contains eleven $z$-by-timing cards.
It also contains seven or eight CTD~FLT
cards which process the $z$-by-timing data from that sector.
The trigger cards dictate the need for custom crates and backplanes 
because of the large number of interconnections required, 360 per slot. 
The performance of the CTD FLT will be the subject of another paper.

A calibration system serves to ensure
the best possible resolution and stability of the $z$-by-timing system and 
also provides a general diagnostic tool to check electrical continuity.
The system includes the following~:

\begin{itemize}

\item{ } Programmable `Calibration Controller' modules
(described in section~\ref{sec:calc}),
one located in each $z$-crate. These
modules supervise a calibration run sequence.

\smallskip 
\item{ } `Calibration Driver' modules 
(described in section~\ref{sec:cald}),
one located in each postamplifier crate. These modules
generate pulses for transmission to the chamber.

\end{itemize}

The readout of the $z$-by-timing and trigger cards
requires modules which control the timing, pipeline addressing,
and local DAQ. The system includes the following~: 

\begin{itemize}

\item{ } Readout Controller (ROC) modules~\cite{ROC1,ROC2}
(described in section~\ref{sec:ROC}), one located in each $z$-crate.
Each ROC supervises the local DAQ of the $z$-by-timing and trigger
data for that crate.

\smallskip
\item{ } A Master Timing Controller (MTC) module
(described in section~\ref{sec:MTC}), housed in an external rack.
The MTC synchronizes the CTD readout with
the overall experimental clock and provides the 
interface to the Global First Level Trigger (GFLT). 

\smallskip
\item{ } Local Timing Controller (LTC) modules 
(described in section~\ref{sec:LTC}), one located in each
$z$-crate.  Each LTC controls the timing and addressing
of the pipelines and buffers of the $z$-by-timing and trigger cards
in that crate. Each LTC and Calibration Controller share a
single 9U~card. 

\end{itemize}

The $r-\phi$~crate contains its own LTC, Calibration Controller and ROC
in addition to the 104~MHz FADC readout cards.

\begin{figure}[htb]
\begin{center}
\epsfig{file=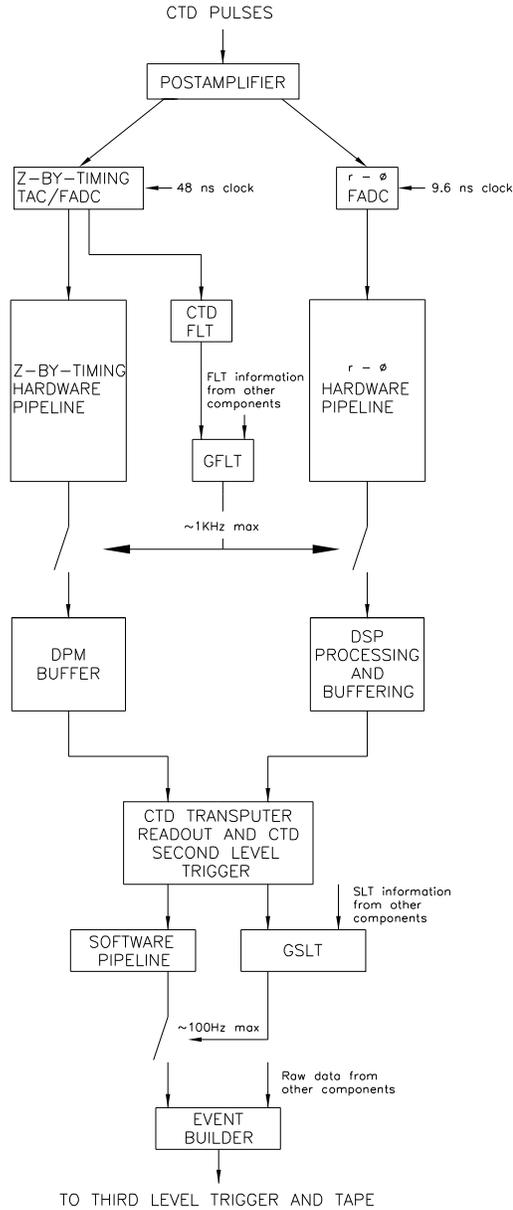,height=16cm}
\caption{ A schematic diagram of the CTD DAQ architecture.}
\vfill
\label{fig:DAQ_overview} 
\end{center}
\end{figure}

A schematic diagram of the CTD readout architecture is shown in 
Fig.~\ref{fig:DAQ_overview}. Triggers at three levels select the 
events and reduce the throughput rate to a maximum of 1~kHz, 100~Hz, and 5~Hz
respectively at each stage.  
Since an FLT decision cannot be made within the HERA
beam crossing time of 96~ns, it is necessary to store the data from all ZEUS
components in local pipelines. Each pipeline contains a record 
of the component data for at least the previous 5~$\mu$s, 
the time to make an FLT decision. Trigger processing is divided between
local component triggers and the GFLT.
The CTD~FLT completes its internal trigger calculation 
after 2~$\mu$s and passes 
information for a particular crossing to the GFLT. The GFLT then 
issues a decision on the strength of the information supplied
from all components. If
no GFLT `Accept' is sent back  to the individual components, the event
data in the pipelines are overwritten.  Events accepted by the GFLT are 
transferred from the local pipelines to the SLT.
As with the FLT, trigger processing at the SLT is divided between 
local component triggers and the Global
Second Level Trigger (GSLT). The CTD~SLT is implemented on
a network of Transputers which run complex trigger algorithms and 
contribute a reduction in the event
rate to a maximum of 100~Hz. Events in this case are stored in local 
software buffers. For events accepted by the GSLT, the digitized 
data from all components are collected together in the Event Builder.
From here events are passed to the
TLT, a processor farm, where for the first time
trigger processing can be performed on the complete
event data from all components. The TLT runs a first-pass 
data analysis algorithm (which includes track finding and fitting).
Events accepted by the TLT are passed to mass storage at a maximum 
rate of 5~Hz.

\section{The CTD Amplification System}
\label{sec:amps}

The amplifiers of the CTD have been described elsewhere~\cite{CTD2},
however since they are a crucial component of the $z$-by-timing
measurement the relevant features are described in this section.

The preamplifiers
are assembled in groups of eight channels onto printed 
circuit boards which form the
preamplifier cards. Each card services a single cell of wires and
is mounted directly onto the chamber endplate.
In this way, each of the 4608 sense wires is connected to a preamplifier
channel at the rear end of the chamber to provide the $r-\phi$ drift-time
measurement. 
To achieve the $z$-by-timing measurement,
the 1152 axial sense wires in Superlayers 1,~3 and 5 have
identical preamplifiers mounted on the forward end of the chamber. 
A total of 704 sense
wires provides hit information to the $z$-by-timing 
readout system; innermost wire~1 to outermost wire~8 in
Superlayer~1 (256 channels) and wires 1,~3,~5 and~7 
in Superlayers~3 and~5 (192 and 256 channels respectively). 

The voltage gain of the preamplifiers is constrained
to a relatively low value, 2.5, in order to minimize
power dissipation (16~mW per channel). 
The first transistor stage of the preamplifier
has a common-base configuration such that the emitter current can 
be tuned to actively terminate the sense wires
in the characteristic impedance of the chamber, 360~$\Omega$.
The need to  minimize pulse reflections caused by improper termination 
at the ends of a wire is well known, and
is crucial to the technique~\cite{zbyt_2,Boie}. These reflections 
result in a non-linear response of
the time difference measurement as a function of the $z$~position of the
hit. As described in~\cite{CTD2,zbyt_2}, 
a mismatch of the characteristic impedance 
is caused by a 7~pF capacitance at the chamber
endplates introduced by the wire mountings,
and the 4~pF input capacitance of the preamplifiers. 
The introduction of a 1$\mu$H inductor at the front-end of the 
preamplifier card forms a $\pi$-section of transmission line
and improves chamber termination and hence linearity.
A measurement of the residual non-linearity is described in 
section~\ref{sec:ZbyT_anRO} and shown in Fig.~\ref{fig:Sshape}(a).

The signals from the preamplifiers are transmitted to postamplifiers
via coaxial cables with total 
lengths of 45.5~m and 42.5~m from the forward and
rear ends of the chamber respectively.
Two types of signal cable are used. 
The preamplifiers are connected 
via lengths of thin 1.6~mm outer diameter (O.D.) foam-core coaxial
cable to the `Services Ring' (see Fig.~\ref{fig:block} and Ref.~\cite{CTD2});
the forward and rear ends are connected by lengths of approximately 3.5~m
and 0.5~m respectively. At the Services Ring
the thin coax is interfaced via bulkhead connectors
to 42~m of high quality foam-core
2.8~mm O.D. coaxial cable, which transmits the signals
to the postamplifiers. Ten cables are used for each
cell of eight wires: eight signals, one power and one calibration. 
The attenuation of the cable at
100~MHz is 46~dB/100m for the 1.6~mm O.D. cable and 22.5~dB/100m for the
2.8~mm O.D. cable. 
High frequency compensation circuitry on the preamplifier
ensures that the preamplifier plus
cable response is flat to $\approx$100~MHz. 

The postamplifiers 
provide an additional voltage gain of 80 for the $z$-by-timing system 
and also split the pulses to the $z$-by-timing and $r-\phi$ FADC systems.
The basic gain for each channel is
provided by an NE592 video amplifier. 

Maintenance of bandwidth throughout the system is crucial for the
$z$-by-timing technique.
The complete amplification system 
(preamplifier, cable and postamplifier) has
a 3~dB bandwidth of 80~MHz. It is also essential to maintain low noise;
the RMS noise amplitude from all sources is
approximately 220~nV/$\sqrt{\rm{Hz}}$ (dominated by the postamplifier).
The amplification system described above
results in average 10--90\% rise-times of 10~ns for chamber pulses  
at the input to the $z$-by-timing readout cards.


\section{The $z$-by-Timing Readout System}
\label{sec:ZbyT}

Signals from the postamplifiers are fed into the $z$-by-timing readout cards
whose function is to digitize the time difference from each wire
into a 7-bit number.
The sensitive analogue circuitry of a single channel
resides on an
analogue `daughter-board' which is equipped to
make the timing measurement and provide its calibration.
Four daughter-board channels are mounted on a 
single `mother-board' which contains the digital readout circuitry.
In this way a single cell of eight wires is serviced by two cards.
The mother-boards are 400~mm deep and
are housed in a 9U high Teradyne crate with a 20~mm pitch. 
Each $z$-crate contains eleven $z$-by-timing cards, 
making 44 channels in total.
The crate also contains seven or eight CTD~FLT modules which
process the $z$-by-timing data from that sector.

\subsection{The Analogue Time Digitization}
\label{sec:ZbyT_anRO}

\begin{figure}[htb]
\begin{center}
\epsfig{file=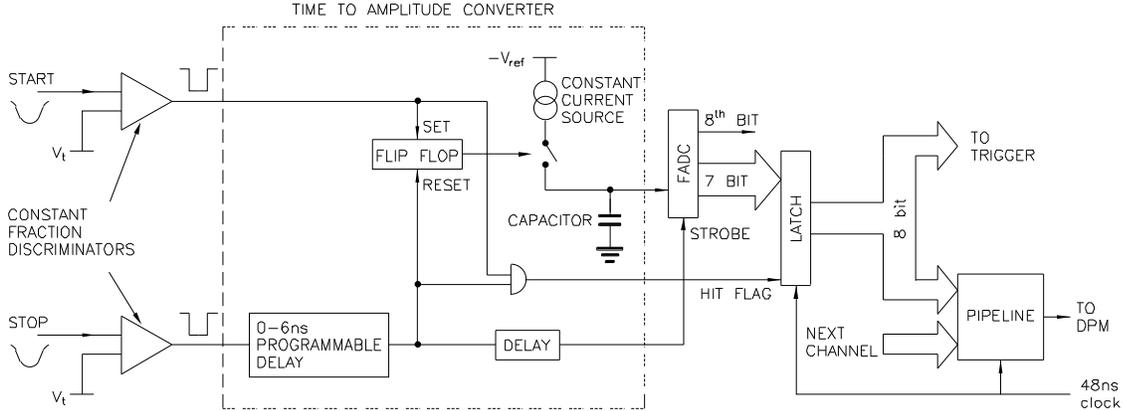,height=5.5cm}
\caption{ A schematic diagram of the front-end $z$-by-timing
readout from Time-to-Amplitude Converter to pipeline. All circuitry, with the 
exception of the pipeline RAM, resides on a single analogue 
daughter-board.}
\vfill
\label{fig:tac} 
\end{center}
\end{figure}

A block diagram of the front-end of a single analogue
readout channel is shown in Fig.~\ref{fig:tac}.
The digitization process can be divided into stages.
At the input of the card, the analogue `start' and `stop' pulses are
sampled by constant fraction discriminators.
The constant fraction of 0.5 and delay of
5~ns have been carefully chosen to provide accurate timing
and to match the input pulse shape characteristics with minimal timing 
slew~\cite{Khatri}.
The discriminated pulse from the rear end of the chamber
provides the start pulse to the TAC,
the pulse from the forward end provides the stop. 
The additional 3~m of cable which runs along the length of the CTD
from its forward end to the Services Ring ensures 
that the start pulse will always arrive before the stop for a valid hit.
There is a programmable digital delay in the stop circuitry
to fine-tune the relative timing (see section~\ref{sec:ZbyT_cal}).

\begin{figure}[htb]
\begin{center}
\epsfig{file=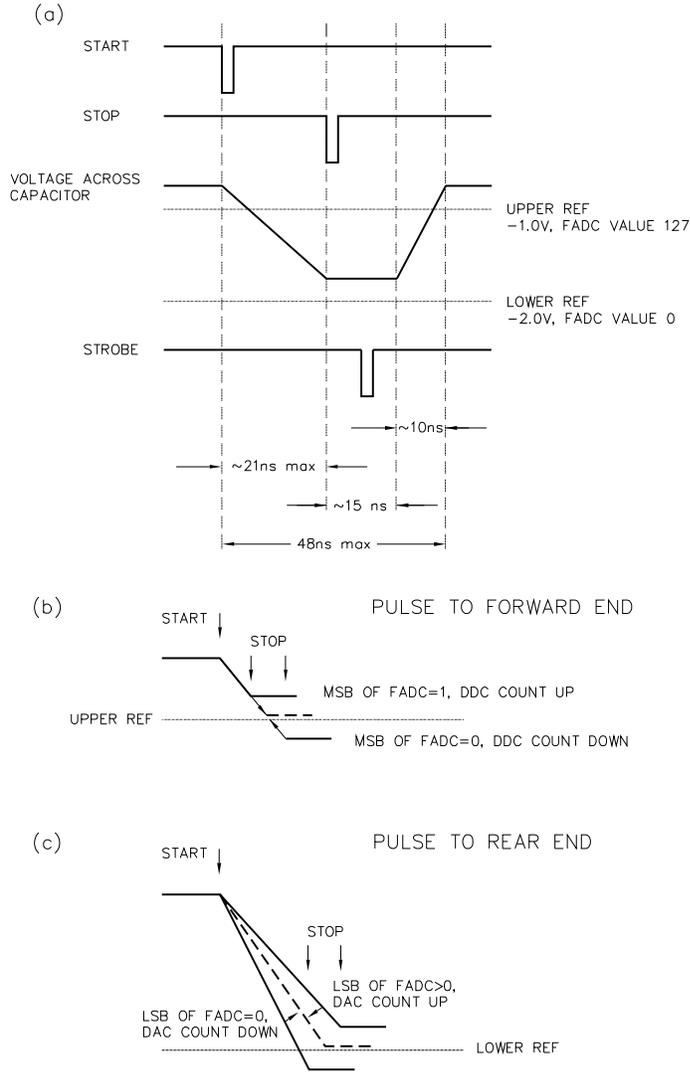,height=15cm}
\caption{ Timing diagram showing the voltage across the 
capacitor : (a) in normal data-taking mode, (b) showing a calibration 
pulse to the forward end of the chamber and (c) showing a calibration 
pulse to the rear end of the chamber. `LSB' and `MSB' are the least and
most significant FADC bits respectively.}
\vfill
\label{fig:cal_seq} 
\end{center}
\end{figure}

The TAC generates a voltage level proportional to the time
difference between the start and stop signals which is then sampled by an
8-bit FADC; a similar technique has previously been used by OPAL~\cite{OPAL}. 
A timing diagram for the conversion is shown in 
Fig.~\ref{fig:cal_seq}(a).
On arrival of a start pulse a capacitor is charged at constant current 
until a stop pulse is received. 
A strobe derived from the stop pulse (suitably
delayed to allow for the maximum ramp time and for the 
settling of the flat top) triggers the sampling of the stored charge.
This sampling is achieved by the 8-bit FADC; only the seven least significant
bits are used to digitize the voltage, the eighth bit being used as an
overflow. The 7-bit output of the 
FADC is subsequently combined with an $8^{th}$~bit validity flag 
(`hit flag') derived from a valid start-stop pair, shown
in Fig.~\ref{fig:tac}.
The 8-bit data are then held in an output latch awaiting a synchronous
copy into a RAM (pipeline) memory.
A conversion will occur only if a start
is followed by a stop within a specified time. 
If a valid stop has not been received within $\sim$40~ns 
after the arrival of a start, there is a `time-out' 
and the ramp is reset ready for the next hit.

The full range of measured time differences is $\sim$16~ns.
This is derived from the travel times of pulses down
the 2.03~m wire length (giving a nominal time difference range
of 13.3~ns), with additional contributions from
small non-linear and end effects.
Seven-bit sampling was chosen so as to give a negligible 
contribution to the overall
resolution arising from the TAC binning.
(Note that the intrinsic timing resolution of the 7-bit TAC is 
$\sim$16~ns/(128$\times\sqrt{\rm{12}})$=36~ps,
equivalent to $\sim$5~mm of wire length.
This represents a negligible contribution to the overall
resolution of $\sim$4.4~cm, the measurement of which
will be described in Section~\ref{sec:results}.)

The time between multihits of the $z$-by-timing measurement is determined
by the maximum time between a valid start-stop combination ($\sim$16~ns),
the sampling time ($\sim$15~ns flat-top), and 
the discharge and re-settling time of the capacitor ($\sim$10~ns).
Added to this is the time 
for the ramp to reach the upper threshold level, chosen
to be 5~ns, which ensures that operation is always
on a constant slope.  This results in
a maximum ramp length of  $\sim$21~ns for any valid start-stop combination.
We have chosen to
clock the system at 48~ns which is double the HERA beam crossing frequency. 
Since the maximum conversion time of a valid hit is always less
than 48~ns, we gate the system so
that a second hit $cannot$ occur within 48~ns of the first.
Although a prompt conversion
could in principle allow the system a faster recovery time, we have chosen
a solution which brings uniformity to all channels for all hits. 
The second
hit resolution of 48~ns nicely matches the equivalent multihit
resolution of the $r-\phi$ FADC system, approximately 50~ns, which
corresponds to a 2.5~mm drift distance.

Every 48~ns, the 8-bit data from the FADC output latch
are clocked into a 8$\times$1k~bit
RAM pipeline memory (details of the clocking procedure
are given in section~\ref{sec:LTC}). 
In the usual scenario of null data from a given wire, zeros will be written
into the pipeline; a zero 
hit flag means that no start-stop combination has been received. 
In the case of a valid hit, the location of the data in the pipeline
memory relative to the beam crossing location 
in which the interaction occurred
provides a coarse measure of the $r-\phi$ drift time.
The error on this drift-time measurement is approximately
48/$\sqrt{\rm{12}}$~ns and, in combination with the digitization
of the $z$~coordinate, 
gives a stand-alone three-dimensional space-point measurement
of a hit. This
is an important feature of the system.

\begin{figure}[htb]
\begin{center}
\epsfig{file=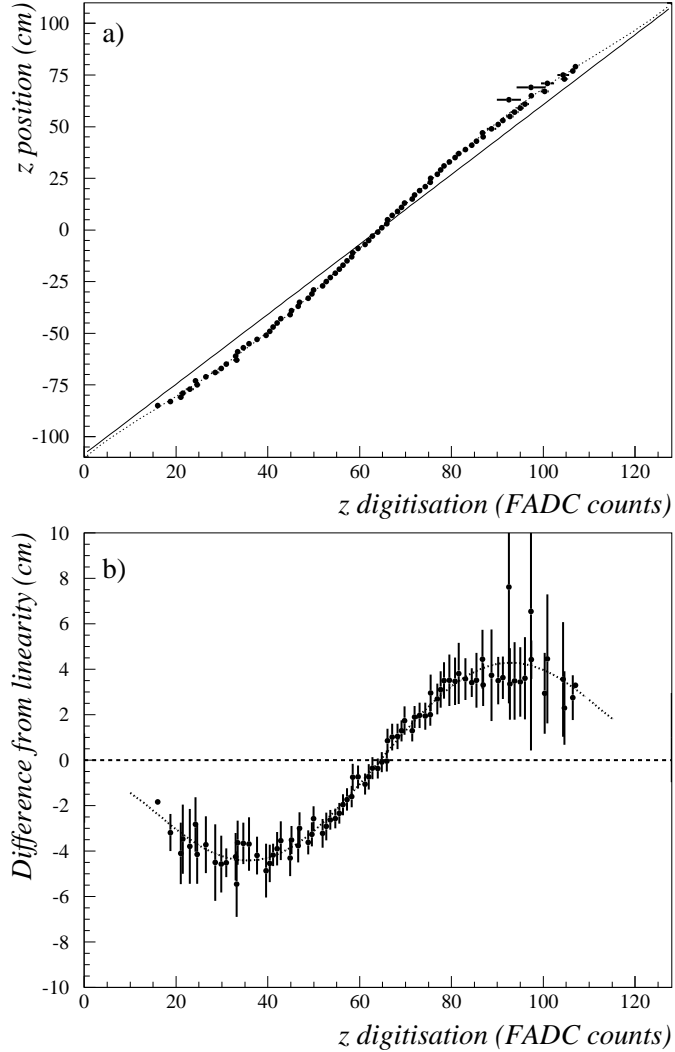,height=14cm}
\caption{
(a) The $z$-by-timing
digitized time difference as a function of reconstructed
$z$~coordinate of the hit, measured
using $e-p$ collision data. The solid line represents linearity,
with the ends of the line constrained to the chamber end-points.
(b) The deviation of the response from linearity.
A fit to these data, represented by the dotted line,  
is described by an antisymmetric quintic function of the
form $z=P_1(t-P_0)+P_2(t-P_0)^3+P_3(t-P_0)^5$. This
allows  a conversion between time $t$ (in FADC counts)
and the $z$~coordinate of the hit (in cm). 
The fit parameters are as follows~: $P_0$=64.2;
$P_1$=2.16, $P_2$=$-$2.30$\times$10$^{-4}$,
$P_3$=$-$2.99$\times$10$^{-8}$~cm. }
\vfill
\label{fig:Sshape}
\end{center}
\end{figure}

The performance of the TAC digitization is demonstrated in 
Fig.~\ref{fig:Sshape}(a). This
shows the distribution of digitized
time difference as a function of the reconstructed $z$~coordinate
of a hit for all instrumented wires, measured
in the ZEUS experiment using tracks from $e-p$ collisions.
The solid line represents the simplest linear form for the response,
constrained by digitizings of 0 and 127 at the wire ends.
As discussed in section~\ref{sec:amps},
imperfect termination of the chamber sense wire results in 
non-linearities in the time-to-distance relationship.
The deviation of the response from linearity is shown
in  Fig.~\ref{fig:Sshape}(b), with a quintic fit superimposed.

The hit data from each FADC output latch are clocked to the 
CTD~FLT~\cite{CTD_trigger}, 
which determines whether a track is consistent
with originating from the interaction region by considering $z/r$ versus $r$
patterns (where $r$ is the radial  position of the 
wire). The conversion from~$r$ to~$z/r$ 
is implemented using a preprogrammed
ROM look-up table, located on the mother-board of each $z$-by-timing card. 
The data are also corrected for non-linearities 
into the FLT by preprogramming the look-up tables 
according to the quintic parameterization of Fig.~\ref{fig:Sshape}(b).

\subsection{Calibration Control}
\label{sec:ZbyT_cal}

Calibration of the $z$-by-timing system is crucial to its
successful operation.
Since we require sub-nanosecond accuracy, the system is extremely sensitive
to small delay changes, e.g. 
temperature fluctuations in cables and electronics.
We have designed the system to calibrate
out such changes either during data-taking or non~data-taking periods.
The logic necessary to perform $z$-by-timing 
calibration is implemented entirely 
in hardware with the only software interaction being the preparatory
configuration of the front-end cards. 
For effective 
operation, a channel is calibrated when the full 7-bit dynamic range of 
the FADC corresponds to the full length of the chamber wire, 
i.e. the extreme FADC values 0 and 127 are output  for tracks which
deposit ionization at the wire end-points.

\begin{figure}[htb]
\begin{center}
\epsfig{file=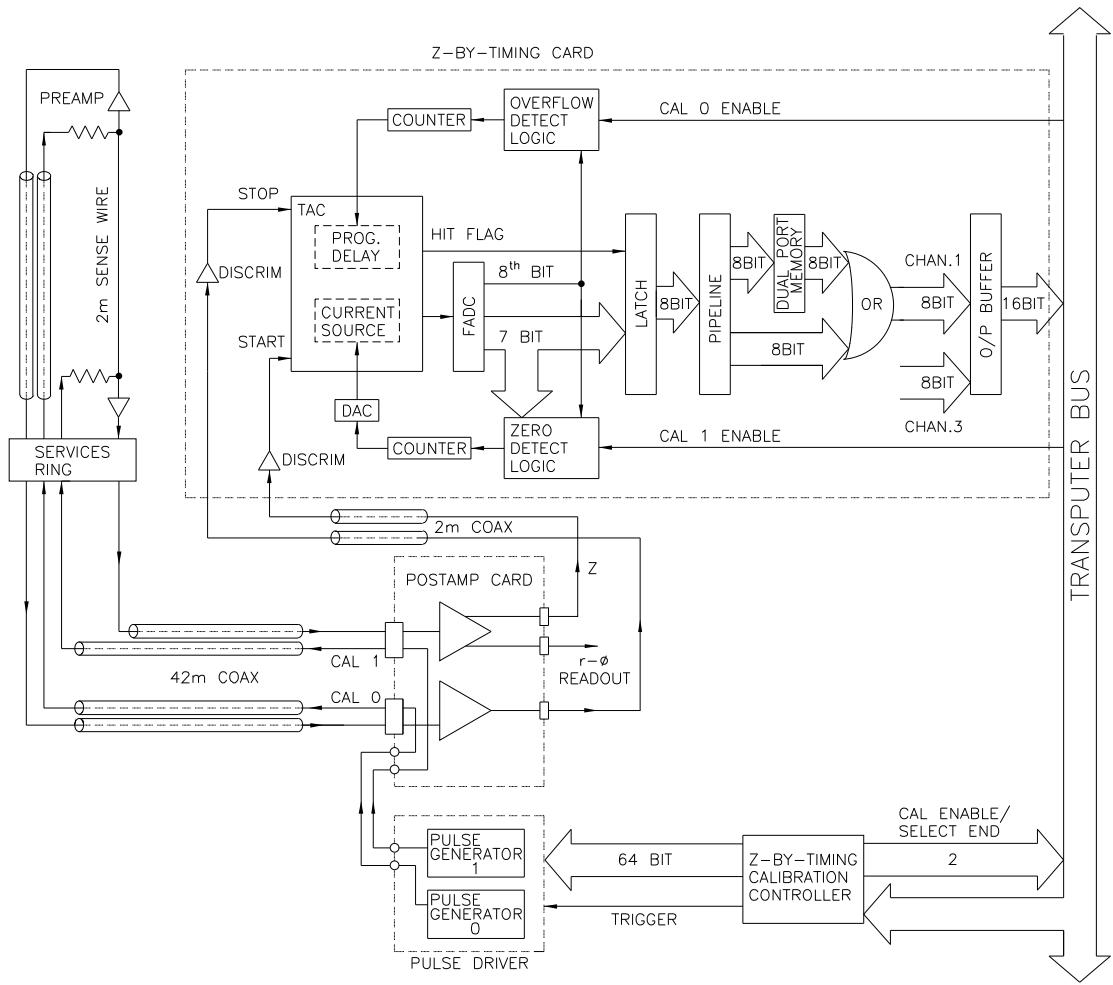,height=13cm}
\caption{ A block diagram showing the calibration control of 
the $z$-by-timing readout system.}
\vfill
\label{fig:cal} 
\end{center}
\end{figure}

A schematic diagram of the components used in the calibration process
is shown in Fig.~\ref{fig:cal}. Timing diagrams for the calibration
sequence are shown in Figs.~\ref{fig:cal_seq}(b) and~(c). 
During calibration, pulses of a size and shape similar to real chamber
pulses are sent alternately to the appropriate
preamplifier cards at either end of the chamber
(details of the pulse injection procedure
will be given in section~\ref{sec:cal}). 
The calibration point of each channel on the preamplifier card 
is in close proximity to the signal input in order
to approximate charge injection from the physical end of the sense wire.
To maintain calibration, two parameters in the TAC are
available for adjustment. The first is the relative arrival of the start and
stop signals, controlled by an 8-bit 0-6~ns programmable Digital
Delay Chip (DDC), which adds delay to the stop line. 
When the forward end of the wire is being pulsed, the value of the DDC
is increased by a single bit if the FADC issues an overflow
(i.e. the 8$^{th}$ bit is set), or decreased 
by a single bit otherwise. 
Hence at the point of
calibration the value issued by the FADC ideally oscillates between 127
and 128 when successive calibration pulses are sent. 
The second parameter is the slope of the constant current 
ramp which charges the capacitor.
When the rear end is being pulsed, an 8-bit Digital to Analogue
Converter (DAC) is incremented
if the FADC issues a value greater
than 0 and this increases the slope of the current ramp. Conversely
if the FADC value equals 0 the DAC is decremented and this decreases the slope.
Hence calibration is ideally marked by oscillation between
no FADC bits set and the first bit set.
The delay and slope parameters are not independent, hence the
need to pulse the ends alternately. 

Each DAC and DDC is controlled by its own 8-bit up/down counter which is
incremented or decremented according to the status of the FADC bits.
Naturally the sensitivity of the DAC and DDC bits have to
be much greater than one equivalent FADC bit;
a transition of 6 and 4~bits of DAC and DDC respectively correspond
to the transition of a single FADC bit.

\begin{figure}[htb]
\begin{center}
\epsfig{file=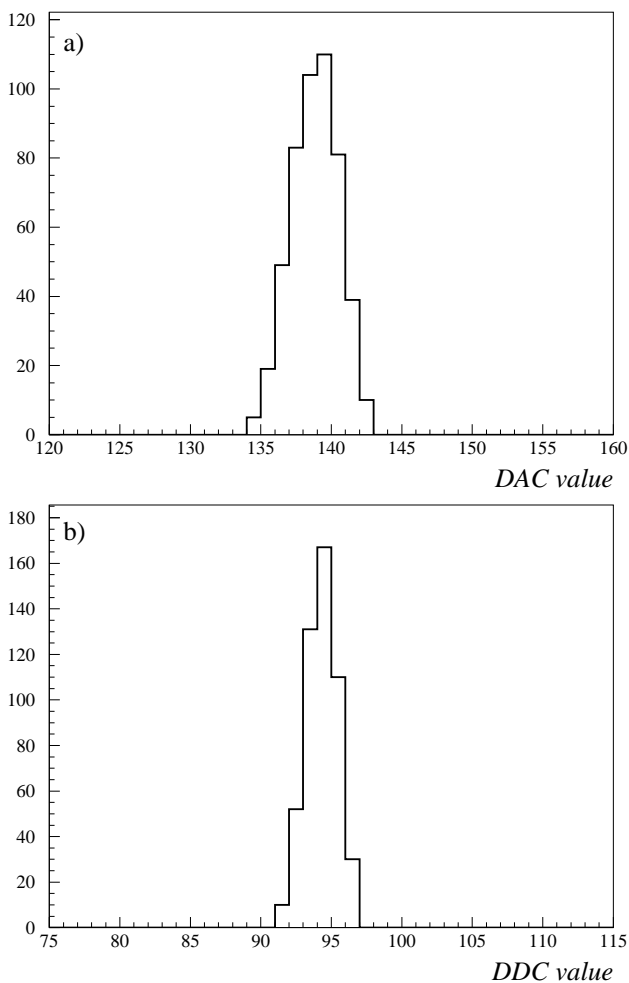,height=13cm}
\caption{
(a) Histogrammed values for typical DAC and (b) DDC counters of a
single channel during a calibration run.}
\vfill
\label{fig:DB_CALIB}
\end{center}
\end{figure}

Generally the $z$-by-timing calibration process is performed 
during non~data-taking periods. 
A calibration `run' is usually made up of a sequence of 
several hundred pulses to each cell of the chamber.
Histogrammed values for typical DAC and DDC counters during calibration 
are shown in Figs.~\ref{fig:DB_CALIB}(a) and (b).
Their mean values are calculated in real time during the calibration run 
and the subsequent values are downloaded to preset the DACs and DDCs
via the backplane bus prior to data-taking.
The standard deviations of the DAC and DDC distributions 
shown in Fig.~\ref{fig:DB_CALIB} are a measure
of the noise and timing instability of that channel.
The average standard deviations over all DACs and DDCs 
in the system are 1.9 and 1.1 counts respectively.

\begin{figure}[htb]
\begin{center}
\epsfig{file=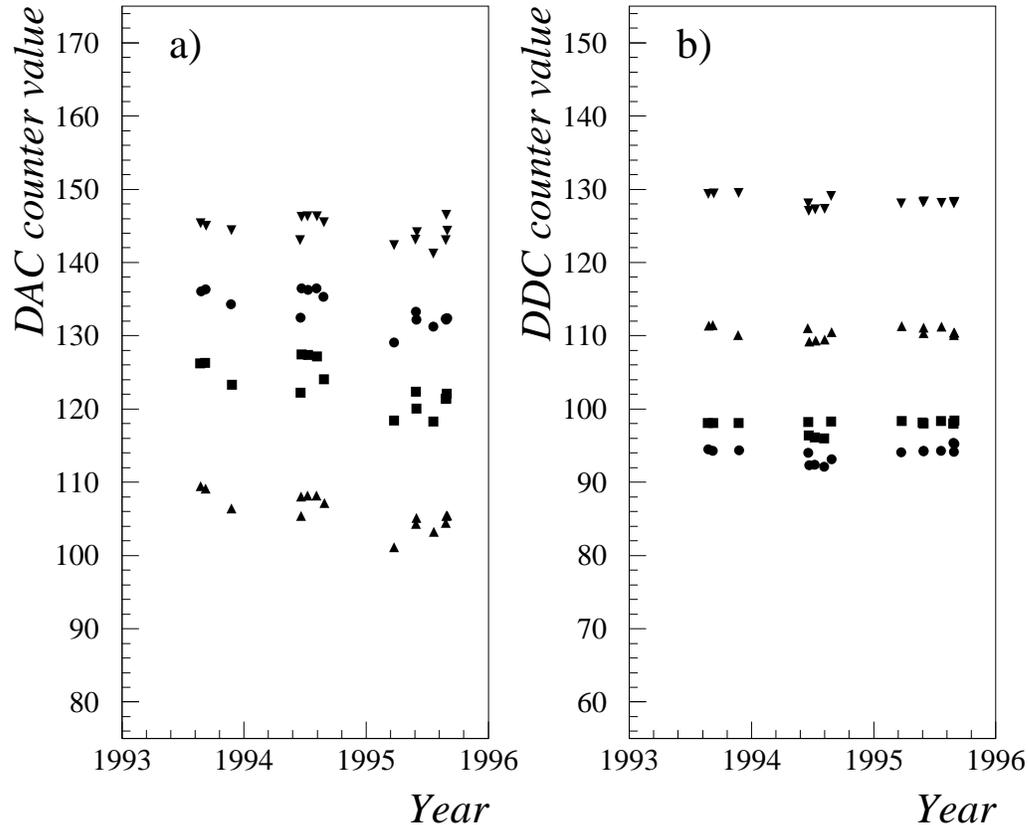,height=11cm}
\caption{
The mean values of (a) the DAC and (b) the DDC 
from the four channels of a typical $z$-by-timing readout card as
a function of time. The data cover a three year period. }
\vfill
\label{fig:DAC_stability} 
\end{center}
\end{figure}

We have found that the time dependence of the DAC and DDC
calibration points  is extremely stable over the course of a HERA
data-taking period. This is demonstrated in 
Figs.~\ref{fig:DAC_stability}~(a) and (b).
Here the mean values of typical DAC and DDC counters are shown as
a function of time, covering a three year period.
We have calculated the mean value of each individual
DAC and DDC counter in the system
over the course of this three year period. Considering how the
DAC and DDC values differ from their mean values over this time
period, we 
subsequently obtain an RMS spread of 3.3 and 1.1 counts 
respectively for
all DAC and DDC counter values in the system.
Given this intrinsic stability, in principle the system 
needs to be recalibrated only infrequently.

\subsection{The Digital Readout}
\label{sec:ZbyT_digRO}

The mother-board contains
four 8-bit RAM pipeline memories (PLMs) which store the data during
FLT processing, four 8-bit dual 
port memories (DPMs) which act as a data store prior to backplane readout,
and all the associated data-transfer logic and backplane interface.
All four channels operate synchronously and in parallel during data-taking.
The pipelines, DPMs and calibration registers are all accessible 
(read and write) from the backplane data bus.
The module can be operated in various modes by 
presetting a function code register (e.g. read or write of
DPM, PLM; select CAL etc).

The 8-bit $z$-by-timing data are clocked from the FADC output latch 
(shown in Fig.~\ref{fig:tac}) into 
sequential locations of the PLM every 48~ns, the 
strobe being derived
from the HERA 96~ns beam-crossing clock.
The length of the pipeline RAM is 1024 locations of memory which
allows data storage
for considerably longer than 5~$\mu$s, the time for the GFLT
to reach a decision. However,
during normal data-taking only the first 440 locations 
of the PLM are used, the 
data being overwritten every 220 beam crossings. This means that the
pipeline address always corresponds to an equivalent HERA bunch crossing 
number; this feature simplifies the DAQ architecture and testing.  
The LTC module, described in 
section~\ref{sec:LTC}, provides the 48~ns clock pulses  and the
appropriate addresses for the PLMs via the crate backplane.

Events which are accepted by the GFLT have their $z$-by-timing data 
transferred from the pipelines into the DPMs, again under control of the LTC
which supplies the appropriate addresses.
During transfer, if a valid hit flag is encountered, all subsequent
hit-flag locations in the DPM for that channel are set high. This speeds
up the DAQ process by testing for valid hit data merely by
reading the $last$ word in the DPM pertaining to that event and channel.
The size of each DPM  is 8$\times$2k~bits which is more than
sufficient to buffer a maximum of 10 events which await readout
(again see section~\ref{sec:LTC}). 

The final stage of the readout is performed by a Transputer-based DAQ
network~\cite{Tariq_paper}. A single Readout Controller in the crate
initiates the transfer of data from all DPMs in that crate
to on-board Transputer memory, and subsequent Second Level processing.


\section{The CTD Calibration System}
\label{sec:cal}

\subsection{Overview}

The CTD calibration system serves a number of purposes :

\begin{itemize}

\item{ } To provide a general diagnostic tool to
check the electrical continuity from 
the end of every instrumented  wire to the readout electronics.
\smallskip
\item{ } To provide shaped pulses with rise-times of $\sim$1~ns
to calibrate the $z$-by-timing system.

\smallskip
\item{ } To provide square
pulses of varying amplitude for $dE/dx$ 
calibration, required to compensate for 
channel-to-channel variation in gain.

\smallskip
\item{ } To provide square or shaped pulses of fixed amplitude
to determine wire-to-wire time offsets ($t_0$'s).

\end{itemize}

The possibility of crosstalk between pulsed cells
means that only a selected pattern of
cells within each sector should be pulsed at a time.
We have therefore
designed flexibility into the system, allowing $any$ combination of
cells to be pulsed independently or simultaneously. 

The components which form the calibration system are as follows~:

\begin{itemize}

\item{ } A `Calibration Controller' module, located in each 
of the $z$~and $r-\phi$~crates. This is a programmable module 
whose purpose is to supervise the calibration sequence and provide timing
synchronization.

\smallskip
\item{ } A `Calibration Driver' module, located in 
each postamplifier crate. This module generates pulses 
of pre-specified amplitude and shape which are then transmitted 
via the postamplifiers to
the chamber. There is an individual pulse generator channel to
serve every preamplifier card.

\end{itemize}

A block diagram showing how the calibration 
modules are interfaced to the $z$-by-timing system
was shown in Fig.~\ref{fig:cal}.

\subsection{The Calibration Controller}
\label{sec:calc}

The calibration sequencing is supervised by 32 Calibration Controller
modules. Two Controller modules service each ${{1}\over{16}}^{th}$
sector; a Controller in the $r-\phi$~crate supervises $dE/dx$
and $t_0$ calibration runs, a Controller in the $z$-crate supervises
the $z$-by-timing calibration.
Each Controller shares a 9U card with a respective LTC.
The two Controllers in the $z$~and $r-\phi$~crates 
corresponding to a particular
azimuthal sector share mastership of that sector's
Driver module; this mastership is mutually exclusive. 

\begin{figure}[tb]
\begin{center}
\epsfig{file=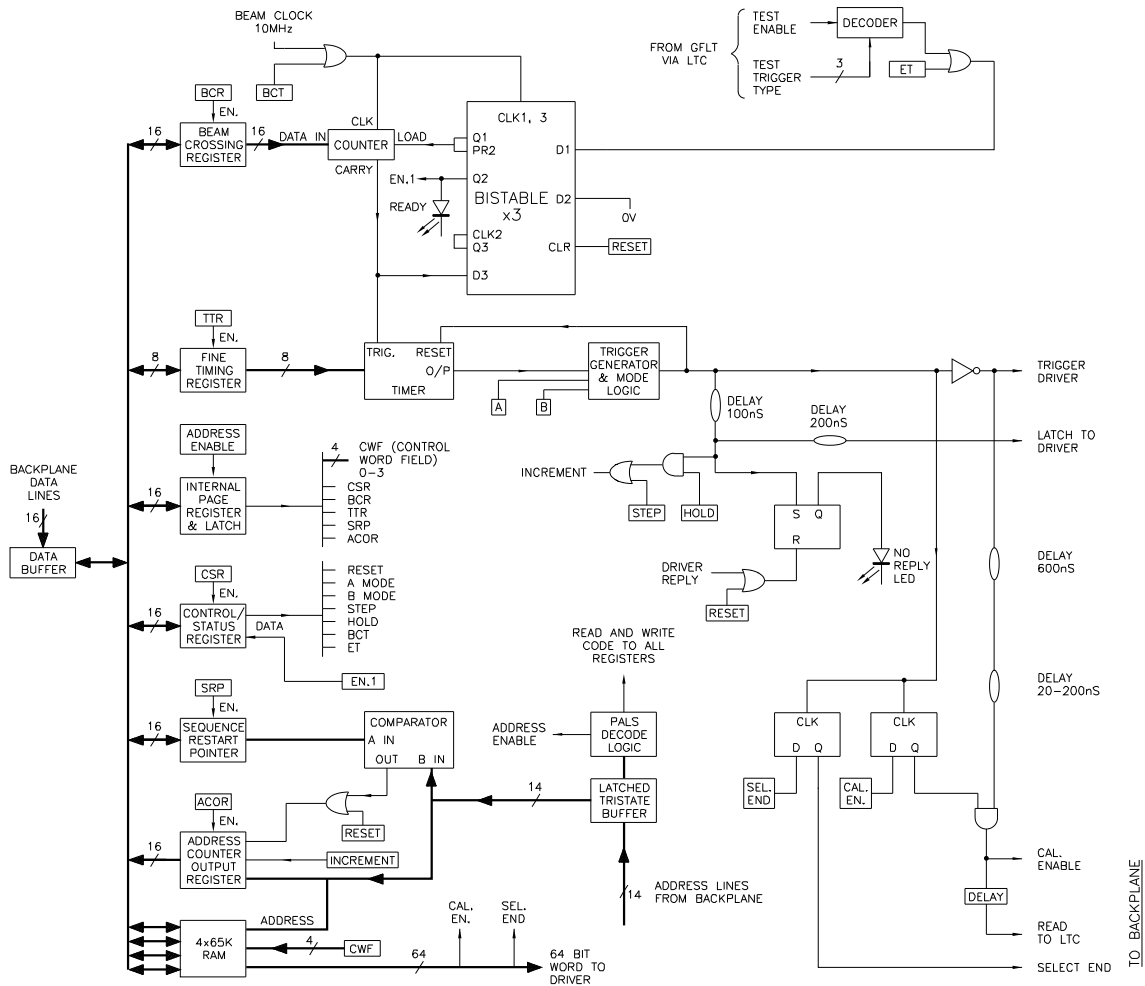,height=13cm}
\caption{ A schematic diagram of the CTD Calibration 
Controller.}
\vfill
\label{fig:cal_control} 
\end{center}
\end{figure}

A schematic diagram of the Calibration Controller 
is shown in Fig.~\ref{fig:cal_control}, to which much of this 
section refers. The Controller
contains its own logic and memory which must be preloaded with a 
calibration `run' definition. 
This is then successively triggered (either under software control
or externally from the GFLT) to enable the specified pulse
sequence. A 64-bit word, loaded in the Controller memory, contains
the bit map of the cells in that sector which will be pulsed simultaneously 
from a single trigger, along with bits to preset
the pulse shape and amplitude. This single pulse-definition
word is latched into the Calibration 
Driver situated in the postamplifier crate of that sector; 
the Driver subsequently  generates the appropriate
calibration pulses on receipt of the trigger from the Controller.
The pulses are then sent to the chamber via the postamplifiers.
When the Driver is triggered and the pulses sent, the next calibration
pulse definition is automatically
downloaded from the Controller to the Driver.
The total size of Controller memory is 64$\times$65k~bits 
consisting of four 16$\times$65k fields, however
only the first 16k locations are used. Hence 16k independent
calibration sequences can 
be sent, which then wrap around on subsequent triggers.
The wrap-around length is set by the `Sequence Restart Pointer' and can
take any value between 2 and 16k.
The Controller uses a page system to communicate with the four
fields of memory; which field
of memory is selected for access by the address bus is controlled by
the `Internal Page Register'. A 4-bit control word (CWF) points
to the required field and this must be preset prior to memory access. 
Control and status of the Calibration Controller
is provided via a 16-bit read/write register, the `Control/Status Register'.
All registers are readable which allows the instantaneous
calibration status to be determined.

During a $z$-by-timing calibration run the 
Controller sends calibration control signals to the $z$-by-timing cards via
the backplane. These signals enable the 
calibration DAC and DDC counting logic (`Cal.~Enable') and
define the end which is being pulsed (`Select End').
The Cal.~Enable signal is a gate which is made wide enough to
accept the full range of arrival times of the analogue
chamber pulses at the input of the $z$-by-timing cards,
but short enough to disable the DAC and DDC counting logic
when arrival times are 
outside this range (i.e. to protect against calibrating
on spurious noise). The relative timing of this
signal is defined by a lumped delay (typically set to
600~ns) on each Controller, which compensates for the return time of
the pulses from the chamber. 

\begin{figure}[htb]
\begin{center}
\epsfig{file=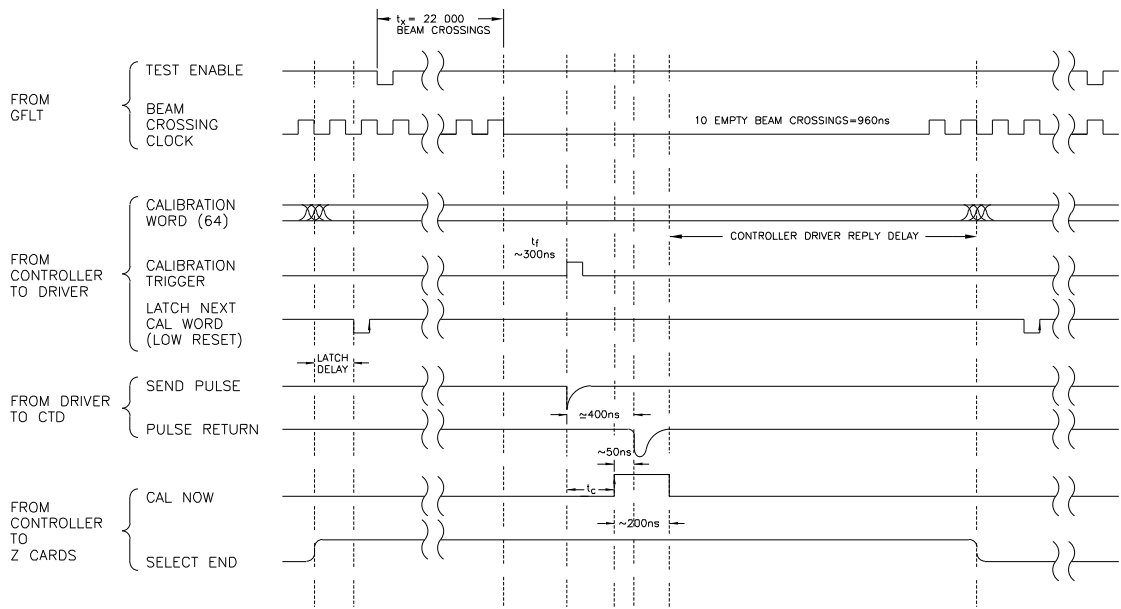,height=8cm}
\caption{ A timing diagram of the CTD Calibration 
Controller sequence during a data-taking run. The delays $t_x$, $t_c$
and $t_f$ are either set in hardware or are programmable.}
\vfill
\label{fig:caltime} 
\end{center}
\end{figure}

Three trigger modes are possible, selected by setting appropriate 
`Mode' bits in the Control/Status Register.
The first mode is `one-shot', used in stand-alone calibration runs. Here
the trigger is under software control and 
is used when calibration is performed prior to a run or during beam-off. 
The second mode is `free-run' (100~Hz), used for diagnostic testing
as well as stand-alone calibration. 
The `Trigger Generator' contains the free running trigger 
generator and the one-shot, as well as the logic to decode the 2-bit Mode
levels from the Control/Status Register. 
The third mode allows calibration during data-taking and here
the overall sequencing of CTD calibration is derived
from the GFLT which initiates a
prescheduled  `Test Enable' sequence, synchronized to the beam clock. 
In this case calibration can occur during the empty beam buckets
of the HERA machine (HERA has 220 bunches, of which 10
consecutive bunches are empty).
The timing diagram for such a sequence is shown in Fig.~\ref{fig:caltime}.
The calibration run definition is downloaded into the
Calibration Controller memory before the run starts.
A total of 220$\times$100
beam crossings before a calibration trigger Accept, the GFLT sends 
a `Test Enable' signal to the CTD. The programmable `Timer'
delay is set such that
when the Controller triggers the Driver and calibration pulses are 
subsequently sent to the chamber, their $z$-by-timing
digitizations are timed to appear in the pipelines just before
the GFLT issues its Accept. 

On the Controller, a 
selectable number of beam-clock cycles (contained in the `Beam Crossing
Register') and a fine adjustment delay (contained in the `Fine Timing
Register') can be preselected to set the delay
between the GFLT Test Enable signal  and Calibration Driver triggering.
The timing of the calibration sequence is controlled by the action of
three bistables.
The Test Enable arrives at D1 of the first bistable
and the next beam clock latches
Q1 high, which loads the contents of the Beam Crossing Register 
into the counter. This register contains the number of beam clocks after 
which a calibration trigger is sent. Q1 also presets the second
bistable (PR2) high, which lights the `Ready' LED.
On the beam clock immediately following
the Test Enable pulse going low, the counter is
enabled to count all subsequent beam-clock pulses. When the specified count
is complete, the carry bit goes high. This then triggers the calibration
sequence after a preprogrammed delay, defined by the Timer
device (which has been previously preloaded from the Fine 
Timing Register). The carry
bit is also input to D3 of the third bistable and through Q2 the
Ready LED is extinguished on the next beam clock. 
The three bistables can be reset from software via the Control/Status
Register.

On receipt of an output pulse from the Timer,
the Trigger Generator sends
a trigger pulse to the Driver as well as a delayed `latch' 
pulse. This latches the memory output into the Driver in preparation
for the $next$ trigger. 
During a  $z$-by-timing calibration run, `Select End' oscillates alternately
high and low on every trigger pulse. 

\subsection{The Calibration Driver}
\label{sec:cald}

A single Calibration Driver module
resides in each postamplifier crate. The function of this
module is to generate pulses which are then sent
passively via the postamplifiers 
down the $\sim$45~m of coaxial cable to the preamplifiers. Charge is
thus injected onto the preamplifier inputs,
as close as possible to the physical ends of the sense wire.
The Driver contains 46 pulse generators, each of which 
can produce a fast shaped pulse or a square pulse. 
Each pulse generator sends pulses to a 
single cell, hence all eight wires of a cell are pulsed
simultaneously. 
During a $z$-by-timing calibration run, fast chamber-like pulses 
are generated and sent to each end of the chamber alternately. 

\begin{figure}[htb]
\begin{center}
\epsfig{file=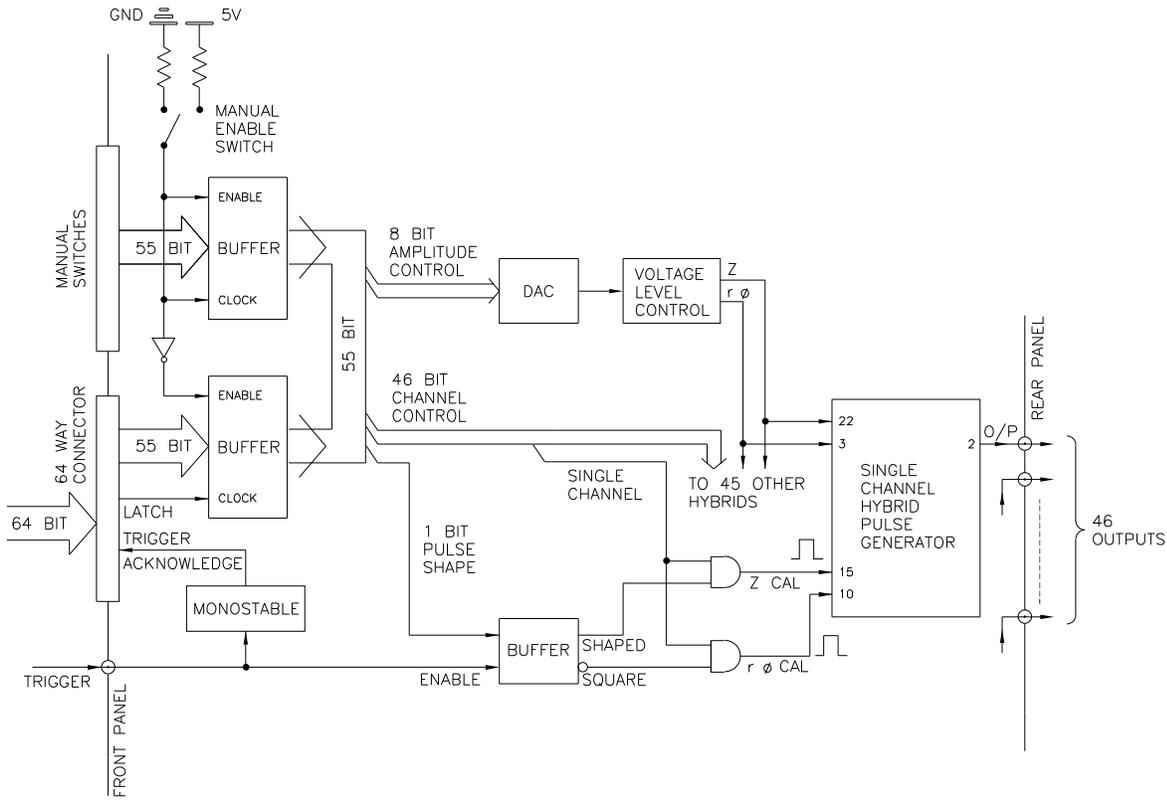,height=10.5cm}
\caption{ A schematic diagram of the CTD Calibration Driver.}
\vfill
\label{fig:cal_driver} 
\end{center}
\end{figure}

A schematic diagram of the Calibration Driver is shown in 
Fig.~\ref{fig:cal_driver}.
The Driver receives the 64-bit control word from the Calibration Controller
and on receipt of a trigger sends out a shaped or square pulse of
a specified amplitude to the preselected channels.
The 64-bit word comprises 46 channel-select bits, one pulse-shape bit, 8
pulse-amplitude bits and additional control bits.
The variation of pulse
amplitude (up to a maximum of 1.2~V) is controlled via 
an 8-bit DAC. 
If required, the Driver can be used independently 
from the Controller by means of front-panel switches.

\begin{figure}[tb]
\begin{center}
\epsfig{file=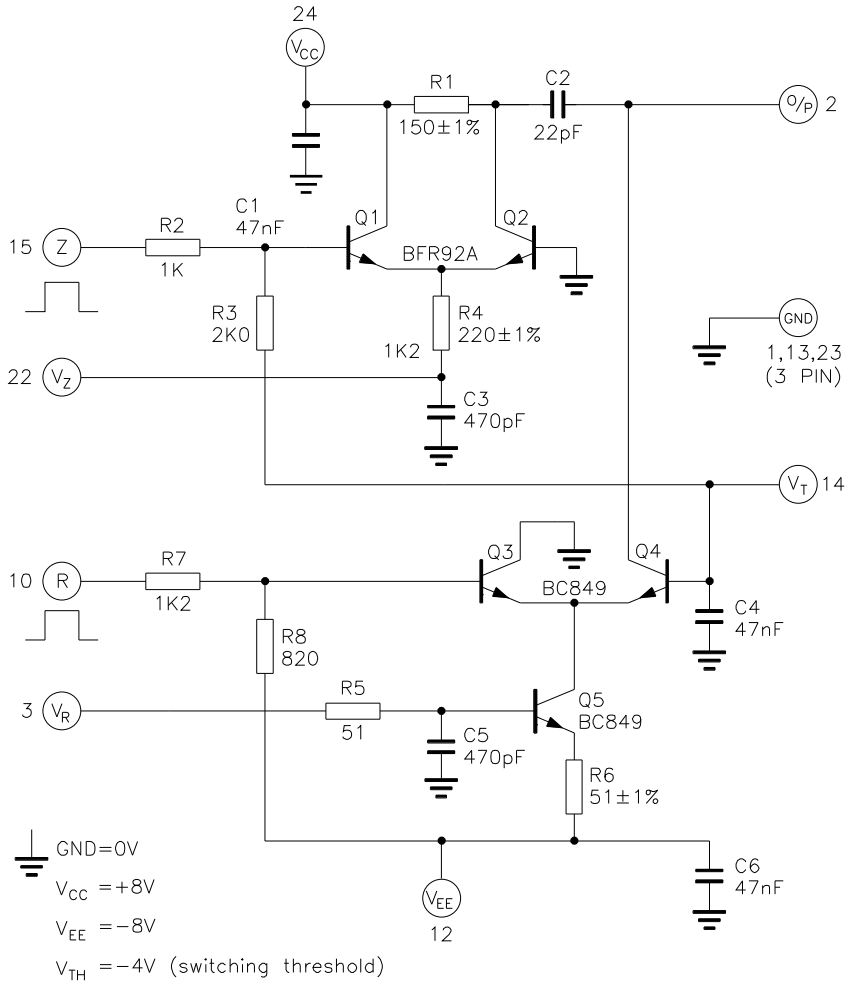,height=13cm}
\caption{ A circuit diagram of one of the
46 Calibration Driver hybrid circuits.}
\vfill
\label{fig:driver_circuit} 
\end{center}
\end{figure}

Each Driver circuit is in the form of an individual hybrid (46 per 
card). A circuit
diagram of the hybrid is shown in Fig.~\ref{fig:driver_circuit}.
Two DC inputs ($V_Z$ and $V_R$), one
for shaped pulses and one for square, determine the ambient current and hence
define the amplitude of the output pulse. The shaped pulse is generated
by switching a differential pair of high-speed transistors from one state
to the other. 
When switching occurs, current flows through Q2 causing a voltage drop
at the collector of Q1; this step-voltage is then differentiated by C2.
This results in a fast output pulse with a rise-time of 1~ns
(dispersion in the 45~m cable increases this to $\sim$2~ns at the 
preamplifier input). The maximum pulse duration is 20~ns (FWHM 5~ns),
the absolute timing jitter is less than 25~ps.
The channel-to-channel amplitude variation is of the order of 5\%.
The square pulse is similarly generated, however here the switching
transition is not differentiated and the negative-going output pulse has
a width defined by the trigger signal (approximately 250~ns) with
a rise-time of 10~ns at the Driver output under no load. 
$RC$ effects considerably
increase the pulse rise-time from the chamber and hence
the trigger pulse has to be chosen wide enough so a flat top is reached.
The channel-to-channel amplitude variation
of the square pulses is controlled within one card
to about 2\%, between cards to 5\%.
The pulses from the hybrid circuits typically have
peak-to-peak noise levels less than 5~mV; those with levels greater
than 10~mV have been rejected.

\section{The $z$-by-Timing Data Acquisition System}
\label{sec:DAQ}

\subsection{Overview}

Local data acquisition of the $z$-by-timing 
data is supervised by a Readout Controller,
one per crate. Its purpose is to read out and zero-suppress
the data, perform calculations on the data (in particular 
SLT processing) and to pass the data onto
the ZEUS Event Builder for subsequent processing.
Timing control is provided by a single Master Timing Controller 
interfaced directly to the GFLT,
and 16 Local Timing Controllers, one per $z$-crate. 

The Timing Controllers have the following functions :

\begin{itemize}

\item{ }  To synchronize data from the 16 $z$-crates.

\smallskip
\item{ }  To interface with the GFLT.

\smallskip
\item{ } To control the pipeline and 
buffers of the front-end electronics.

\smallskip
\item{ }  To interface with the DAQ software.

\end{itemize}

\begin{figure}[htb]
\begin{center}
\epsfig{file=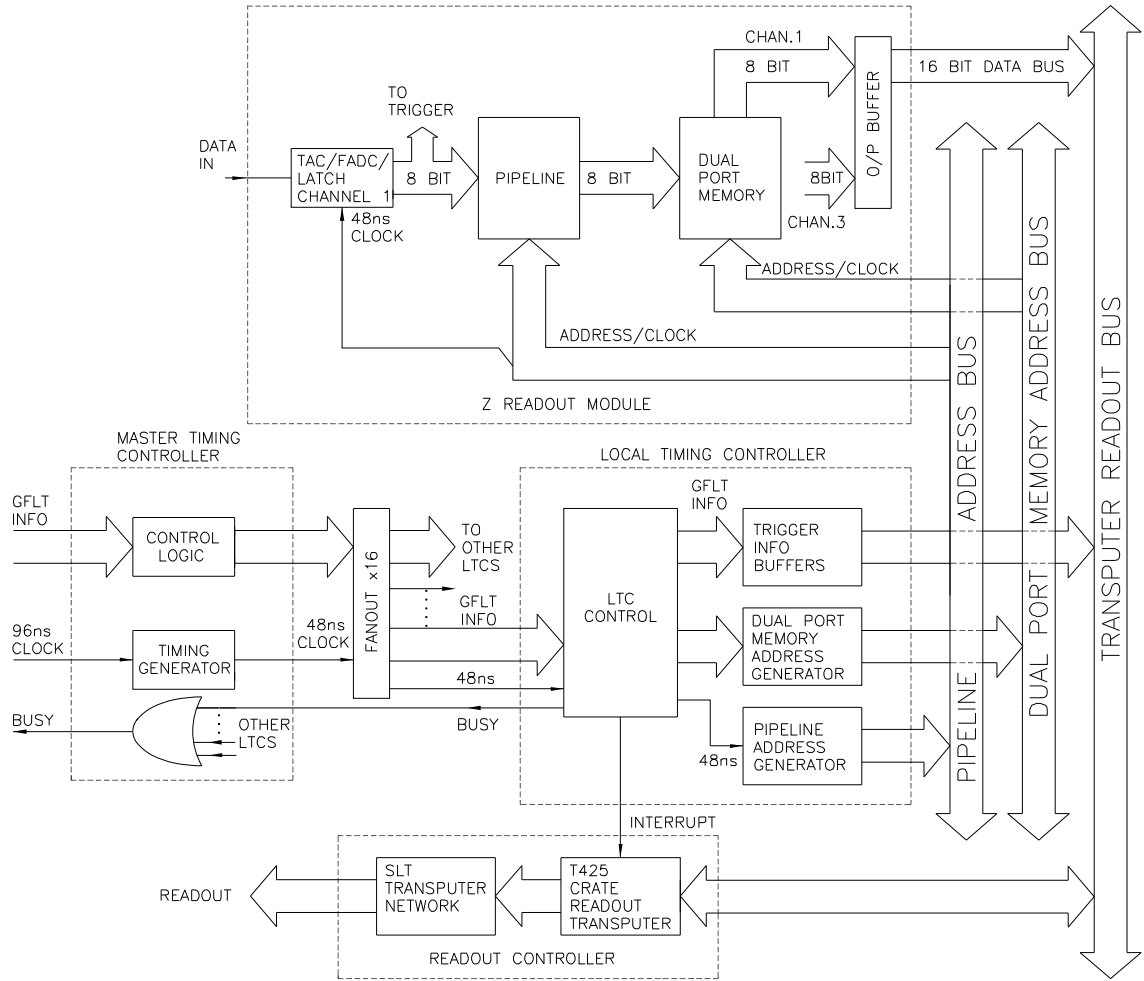,height=13cm}
\caption{ The local $z$-by-timing DAQ system,
demonstrating the functionality of the Timing Controllers.}
\vfill
\label{fig:DAQsys} 
\end{center}
\end{figure}

Fig.~\ref{fig:DAQsys} is a schematic diagram showing the functionality
of the timing system and readout control.
The Master and Local Timing Controllers provide timing to both the $z$
and the CTD~FLT systems. The design philosophy of MTC/LTC
communication is one of  simplicity
and robustness; all timing is derived from a single clock and the only 
control signal returning from the LTC to the MTC is a `Busy'.

\subsection{The Readout Controller}
\label{sec:ROC}

The CTD DAQ and SLT
processing system is based on Transputer-based ROCs~\cite{ROC1,ROC2},
each of which controls a single crate. 
The $z$-crate ROC is a single 9U card which 
contains two T425 Transputers with 
a total of 3~Mbytes
of external memory and three `TRAM' modules, each containing a
T800 Transputer with 1~Mbyte of external memory~\cite{Transputer}. 
The ROC operates a backplane bus to download parameters to 
and read data from the front-end cards.
Each Transputer is equipped with a 10~MIPS processor, 
on-chip memory, and four bidirectional 20~Mbit/s serial links which
make it ideal for the application of SLT parallel processing.
In response to an interrupt from the LTC, readout of the crates is 
carried out by one of the T425 Transputers on each ROC, known as the
`Readout Transputer'. The Teradyne backplane is mapped directly
onto the Readout Transputer memory, hence
one can download and read data from 
each of the front-end cards by simply
reading from and writing to locations in memory.
The second T425 performs data merging and transport
to the Event Builder.
The three T800 Transputers are dedicated to running the software-based 
SLT algorithm~\cite{CTD_SLT,Tariq_paper}.
The interface between the front-end $z$-by-timing
cards, the LTC, and the Readout Transputer via the address and
Transputer buses of the crate backplane is shown in Fig.~\ref{fig:DAQsys}.

\subsection{The Master Timing Controller}
\label{sec:MTC}

The main purpose of the MTC is to
synchronize the CTD readout and FLT processing to the 96~ns HERA 
beam-crossing clock (and hence to the rest of the experiment),
and to provide the interface
by which the DAQ system is able to recognize a trigger
Accept. 
The MTC resides in a dedicated crate (the so-called `RBOX' Crate)
which also houses MTC fanout units and
the cards responsible for the final stages of the
first level track trigger processing and the interface to the GFLT.
Although its operation during normal data-taking is relatively 
simple, its test capabilities make the circuitry rather complex.

\begin{figure}[tb]
\begin{center}
\epsfig{file=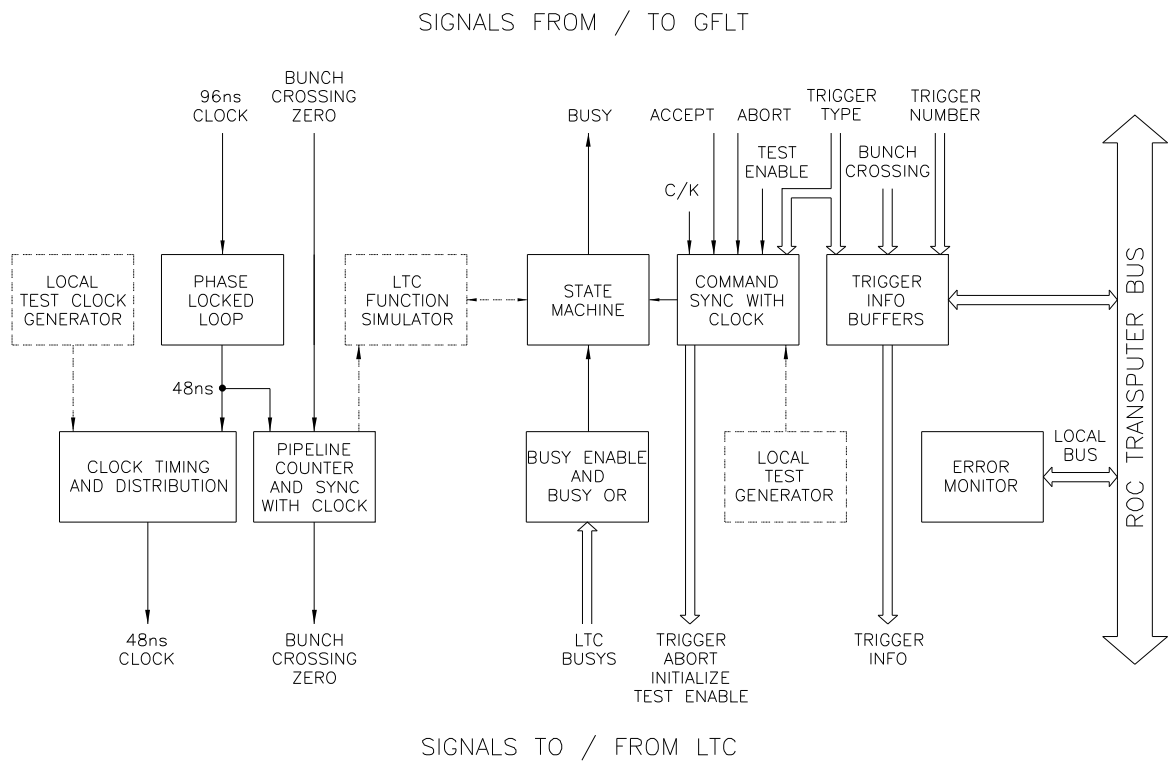,height=10cm}
\caption{
A functional diagram of the MTC.
The test functionality, not operational during normal data-taking, is
represented by dotted lines.}
\vfill
\label{fig:MTC} 
\end{center}
\end{figure}

A functional diagram of the MTC is shown in Fig.~\ref{fig:MTC}.
The MTC receives the 96~ns beam-crossing clock from the GFLT,
derives a 48~ns clock using a phase-locked loop,
and fans out this clock to each of the 16 LTCs.
In addition to the clock, the MTC receives a number of signals from the GFLT:

\begin{itemize}

\item{ }  `Beam Crossing Zero' (BCN0) is a signal which defines
which clock pulse corresponds to the beam crossing numbered zero.
Its role is to synchronize the pipelines throughout
the system by resetting the LTC pipeline addresses to zero 
(see section~\ref{sec:LTC}). Timing 
accuracy of this signal is important, hence
the MTC synchronizes this signal to the 48~ns clock before it is fanned
out to the 16 LTCs.
Although the pipeline address generation is done on the LTCs,
there is also a pipeline counter on the MTC 
which serves two important functions.
Firstly it enables the MTC to detect a `pipeline jump' error if the
BCN0 signal from the GFLT goes out of synchronization
(in which case the MTC will flag an error bit). Secondly,
when running in stand-alone test mode,
it enables the MTC to generate its own BCN0 signal
which is then sent to the LTCs.

\smallskip
\item{ }  `Accept' and `Abort' are the signals that control the system.
The GFLT Accept signifies an event trigger and hence initiates a
pipeline copy. The Abort initiates an abort of the
pipeline copy and on receipt of this signal the system becomes free to restart
clocking data into the pipelines.
The MTC decodes and time-synchronizes these signals with the
48~ns clock before
passing them on to the LTCs. Likewise the Test Enable signal passes
through the MTC and thence to the LTCs and Calibration Controllers.
This signal  instigates a calibration cycle.

\smallskip
\item{ }  The `Bunch Crossing Number', the `Trigger (event) Number' 
and `Trigger Type' signals
are passively transferred from the MTC to the LTC. 
Each LTC subsequently stores this information in DPMs 
which are read out by the ROC.
In this way the LTC tags the relevant $z$-by-timing card buffer with the
Trigger Number and so synchronizes this block of data with
the rest of the experiment.
The Trigger Type contains information such as Initialize, Test Trigger, 
End of Run etc, which is accessed by the MTC and the ROCs.
 
\end{itemize}

When the GFLT issues an Accept, the MTC passes on 
all trigger information to each LTC. 
At any given time
the `State Machine' logic defines what state the system as a whole is in.
Accept, Abort and LTC Busy are the signals which change this state.
The MTC OR's 16 Busy signals, one
from each LTC, the output of  
which is sent back via the State Machine logic to the GFLT.
This Busy, OR'ed at the GFLT with Busy's from other ZEUS
components, inhibits further data-taking until
all Busy's are low.

The MTC provides essential stand-alone test functionality, both self-test
and test of the LTCs. This test circuitry accounts for much of 
the MTC complexity. This capability provides the important
GFLT functions and gives essential diagnostics
in case of failure and during system installation.
During data-taking, the MTC has circuits to detect various errors  such as
clock, address and synchronization faults, signalled by LEDs.

\subsection{The Local Timing Controller}
\label{sec:LTC}

\begin{figure}[tb]
\begin{center}
\epsfig{file=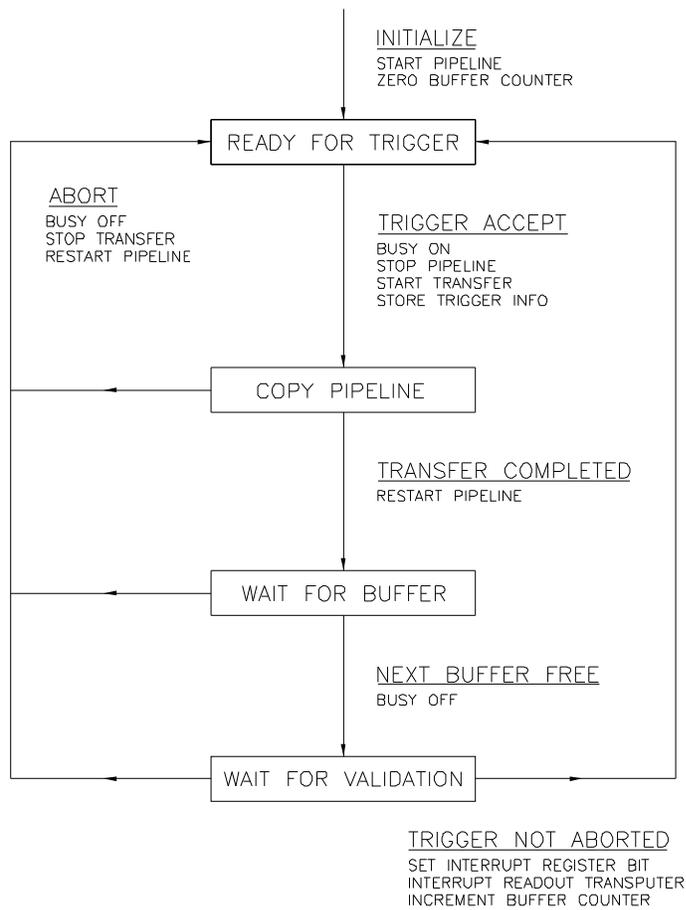,height=12cm}
\caption{
A state transition diagram of the LTC, demonstrating its basic operation
during a run.}
\vfill
\label{fig:STD} 
\end{center}
\end{figure}

The main purpose of the LTCs is to control the pipelines
and buffers of the front-end $z$-by-timing cards, and to interface
with the DAQ system. In addition the LTCs 
clock $z$-by-timing data into and through the trigger processors.
Each of the 16 $z$-crates contains one LTC; the LTC 
shares a 9U card with that crate's Calibration Controller module.
The role of the LTC is demonstrated in
Fig.~\ref{fig:STD} which shows a state transition diagram of its readout
sequence. A description of how the state transitions are implemented in
hardware is given below.

\begin{figure}[htb]
\begin{center}
\epsfig{file=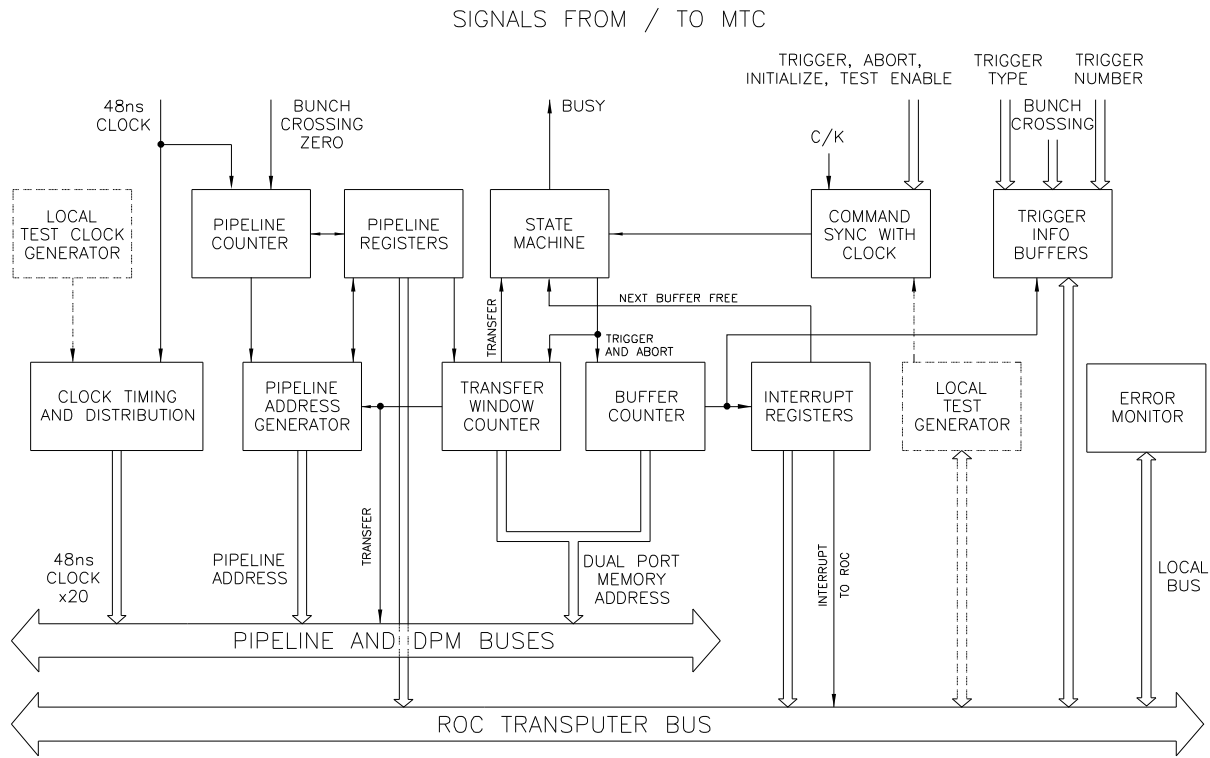,height=10cm}
\caption{
A functional diagram of the LTC. 
The test functionality, not operational during normal data-taking, is
represented by dotted lines.}
\vfill
\label{fig:LTC} 
\end{center}
\end{figure}

A functional diagram of the LTC is shown in Fig.~\ref{fig:LTC}.
The LTCs receive the MTC clock pulses and use them to synchronously generate 
pipeline addresses; this process is 
carried out in parallel across all crates.
The `Pipeline Address Generator' provides the pipeline addresses
to the $z$-by-timing and trigger cards via the crate backplane.
While the data are being pipelined these
addresses are derived from the `Pipeline Counter',
which is incremented every 48~ns  (twice every beam crossing). 
The Pipeline Counter is reset to zero by Bunch Crossing
Zero every 440 addresses (220 bunch crossings).
Prior to a GFLT Accept being received,
the data in the pipeline are simply overwritten by subsequent events
after $\sim$21~$\mu$s.
On receipt of the Accept, a window of data is transferred from
the pipeline memory on each $z$-by-timing card
into the corresponding DPM; the pipeline window addresses are 
provided by the Pipeline Address Generator. The DPM addresses
are also derived from the LTC. During the transfer,
the clocking of new data into the pipeline must be disabled,
hence the system incurs deadtime (typically 1~$\mu$s per
transfer for the $z$-by-timing 
system only, but 3~$\mu$s including the CTD~FLT).  
Throughout the transfer procedure
the Pipeline Counter keeps counting in
order to maintain bunch crossing number integrity.

The interface between the
Pipeline Address Generator and the software running on the ROC
is via the `Pipeline Registers', which are preloaded from the ROC
prior to data-taking. 
Three registers control the pipeline length (normally set to 440 locations),
the length of the window to be read out, and the window `Jumpback'
value (the position of the pipeline window to be
read out relative to the arrival time of the Trigger Accept). 
The window length is 
approximately 50 locations in normal data-taking mode, 
i.e. 2.5~$\mu$s equivalent. The window has to be relatively wide since
the $z$-by-timing and trigger cards have their data in a spread of
pipeline locations because of
the finite processing time through the CTD trigger.
The Jumpback is normally set to 105 locations to account for the timing of
the GFLT Accept which occurs 46 bunch crossings after the interaction.
This defines the start of pipeline readout, 12 
time bins before the interaction crossing.
An additional three registers store pipeline address information 
and include the `Trigger Address', i.e. the
pipeline address at which the last trigger was received.

\begin{figure}[htb]
\begin{center}
\epsfig{file=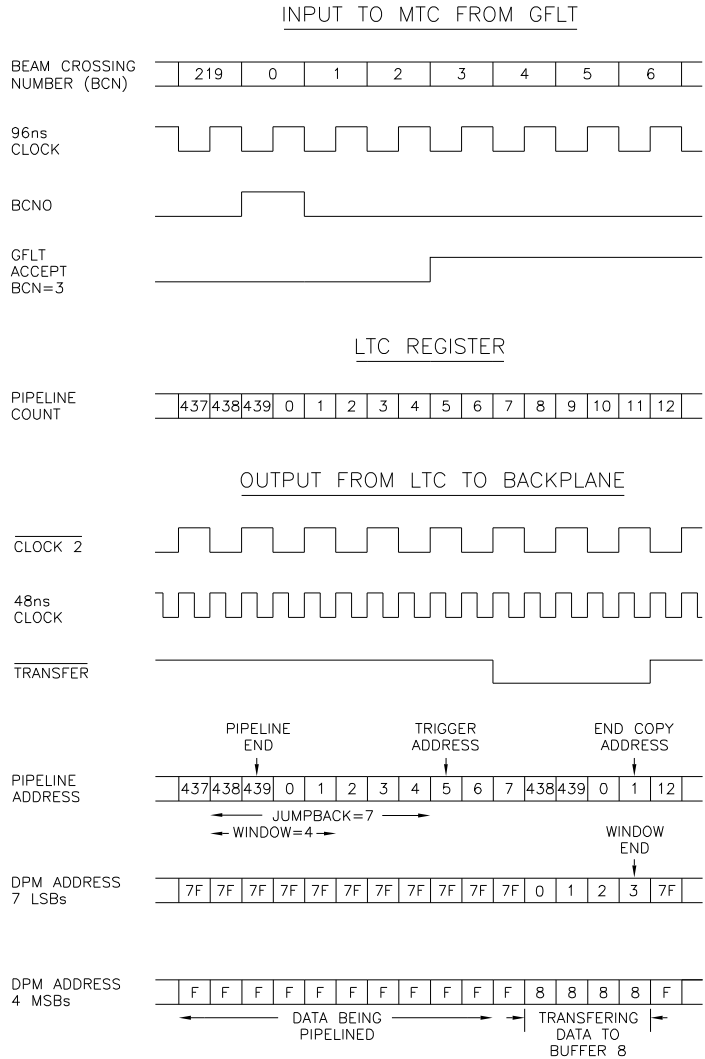,height=14cm}
\caption{
An example of a timing diagram also showing LTC
address transitions. Note that for clarity of the diagram, this
example uses non-typical values of Jumpback and window length
values (7 and 4 respectively). The Buffer Count value has been 
chosen arbitrarily to 8.}
\vfill
\label{fig:LTC_timing} 
\end{center}
\end{figure}

The  `Transfer Window Counter' generates the seven least significant 
bits of the DPM Address during the transfer of data.
It starts counting from zero after a trigger has been received
and signals the end of the transfer when the count reaches
the `Window End Address'.
When asserted, the `Transfer' control signal enables the transfer of data
from pipeline memory to DPM buffer (and is high
for one additional clock period to allow the front-end cards time
to change their memory read/write enables).
The Transfer Window Counter and the Pipeline Address Generator
change mode simultaneously from writing pipeline data
to transferring pipeline data.
The DPM is partitioned into ten `event buffers'
in order to reduce system deadtime.
This means that a maximum
of ten events can be stored for readout by the ROC.
The buffer number, simply the four most significant bits 
of the DPM address, 
is counted by the `Buffer Counter'.
An example of a timing diagram which also demonstrates the relevant LTC
address transitions is shown in Fig.~\ref{fig:LTC_timing}.

The transfer of data from the DPM to the ROC is an asynchronous process.
On command from the LTC,
data are transferred from the DPM to the ROC (and hence to the SLT)
under control of software running on  the ROC Readout Transputer.
The window of data transferred 
in normal data-taking mode is approximately 16 locations, 
i.e. $\sim$800~ns equivalent. 
This is chosen to minimize the overall data volume
through the system while allowing the complete chamber drift time
($\sim$500~ns) to be sampled. 
Interaction with the DAQ software running on
the ROC occurs through an `Interrupt Register' on the LTC. 
Each DPM event buffer is associated with a bit in the Interrupt
Register, which is set from
the Buffer Counter if that event buffer is awaiting readout. 
When an event is validated (i.e. the Trigger Accept is not aborted
by the GFLT), the LTC sets the bit.
In usual operation this is synchronous between all LTCs in the system. 
Any set bit results in an interrupt in the Readout Transputer,
thereby notifying the
DAQ software that an event is awaiting readout from the buffers.
After the appropriate buffers on all $z$-by-timing 
cards have been read out, the 
software clears the Interrupt Register bit for that event, which
frees the buffer. This is asynchronous between LTCs.
Each buffered event is tagged with its Trigger Number as determined
by the GFLT, which later allows
data from all 16~LTCs to be re-synchronized.

Each LTC provides a Busy signal to
the MTC until its pipeline copy is complete. 
The MTC presents a $z$-by-timing system
Busy flag to the GFLT until all 16 LTC Busy's are lowered
(note however that during this period
the MTC continues to provide 48~ns clock pulses
for the trigger processors). 
Additional deadtime will be incurred if any event buffer is full with ten 
events, with the LTC again maintaining its Busy signal.
The LTC `Next Buffer Free' control signal lowers the Busy
via the LTC `State Machine' logic.

Ideally for a given trigger, pipeline addresses 
should be the same throughout all 16 LTCs. 
The window location transferred from pipeline to buffer memory
is determined by the arrival time of the GFLT Accept
at the LTC Pipeline Address Generators. 
Similarly the setting to zero of the pipeline address
is determined by the arrival time of Bunch Crossing Zero.
Throughout the system, the relative $t_0$'s need to be
well controlled. 
The $t_0$'s are affected by the following:

\begin{itemize}

\item{ }  The `Jumpback' values loaded in the Pipeline 
Registers. These are usually the same throughout all crates.

\smallskip
\item{ }  The timing of the HERA/GFLT 96~ns 
clock relative to the interaction time.
This is affected by the details of the tuning of the HERA
machine parameters.

\smallskip
\item{ }  The delay from the GFLT to the MTC.

\smallskip
\item{ }  The timing of the 48~ns 
clock from the MTC through the fanout and LTCs
to the pipeline on the $z$-by-timing cards. This is controlled to about
2~ns across all crates.

\smallskip
\item{ }  The cable and electronic delays 
from the CTD to the pipeline on the $z$-by-timing cards.

\end{itemize}

\noindent The RMS spread in wire-to-wire $t_0$'s from chamber to readout
is approximately 3.6~ns. The spread from clocks on 
the backplane is about 1~ns.


\section{Performance of the System}
\label{sec:results}

The performance of the $z$-by-timing system has been studied using
data collected during the 1994 running
period of the HERA accelerator. 
A total of 180 positron (27.5~GeV) and proton 
(820~GeV) bunches were filled, resulting in 
beam currents of typically 20-30~mA and 30-40~mA respectively. The RMS 
length of the interaction region was approximately 10~cm, dominated by
the proton bunch length. 
Throughout the running period, ZEUS operated with a reduced 
magnetic field of 1.43~T. This necessitated operation of the CTD with an
argon/CO$_2$/ethane gas mix in the
proportions 85/8/7 with a 0.84\% admixture of ethanol
(see Table~\ref{table:gases}). An applied drift field
of 1.22~kV/cm gave a
Lorentz angle and drift velocity close to 45$\degree$ 
and 50~microns/ns respectively.
The sense wire surface fields were chosen to give a gas gain of approximately
1$\times$10$^5$. Operation of the chamber
at relatively high gas gain is crucial for achieving adequate 
$z$-by-timing performance.
        
Throughout the HERA running period,
the $z$-by-timing system
has been extremely stable and reliable. Dead channels have generally
been associated with cable faults. We operate with typically
10 dead channels out of 704, i.e. less than 2\%.

Two independent data-sets have been used for the analysis described
in this section.
The first data-set (the `DIS sample')
uses the standard ZEUS 1994 deep inelastic scattering
selection criteria~\cite{DIS_paper}
which requires an isolated positron candidate measured in the calorimeter
with $Q^2>$5~GeV$^2$.
The second (the `$\rho^0$ sample') uses data which have been tightly
selected as quasi-elastic, diffractively produced
$\rho^0$'s~\cite{rhopaper}. Events in this
sample are extremely clean~-- the ZEUS detector contains two (and only two) 
pion tracks from the $\rho^0$ decay. Unless explicitly stated otherwise,
the data presented below are taken from the $\rho^0$ sample.

\subsection{Hit Matching of $z$-by-Timing and $r-\phi$ FADC Systems}

As described in Section~\ref{sec:CTD_RO}, 
the CTD is instrumented by independent $z$-by-timing
and $r-\phi$ FADC  readout systems (a block diagram for the readout
of a single sense wire was shown in Fig.~\ref{fig:block}).
For the 704 wires instrumented with $z$-by-timing readout,
a match can be made 
between the raw hit information from the two systems. This turns out to be a
powerful method, independent of track reconstruction, for studying
the performance of the chamber and its readout. Both readout systems
measure pulse drift times.
The $r-\phi$ FADC system samples pulses and digitizes voltage every
9.6~ns. Digital Signal Processors (DSPs) on the front-end readout cards 
then compute the drift time to within 2.4~ns and also
the pulse amplitude of each hit~\cite{FADC}.
The $z$-by-timing system digitizes the drift time
in 48~ns bins, as well as giving the $z$~position along the sense
wire from time difference.

\begin{figure}[htb]
\begin{center}
\epsfig{file=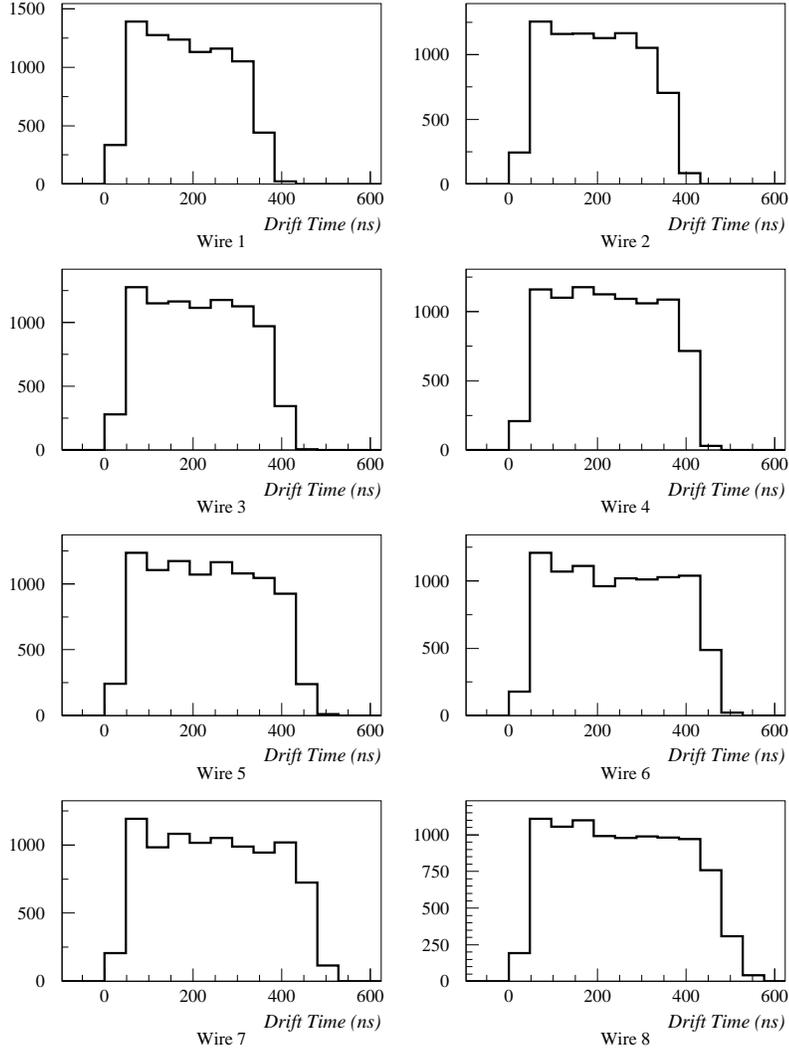,height=14cm}
\caption{
The drift time of hits in bins of 48~ns measured by the $z$-by-timing system
for the eight wire-layers of Superlayer~1.}
\vfill
\label{fig:drift_time}
\end{center}
\end{figure}

\begin{figure}[htb]
\begin{center}
\epsfig{file=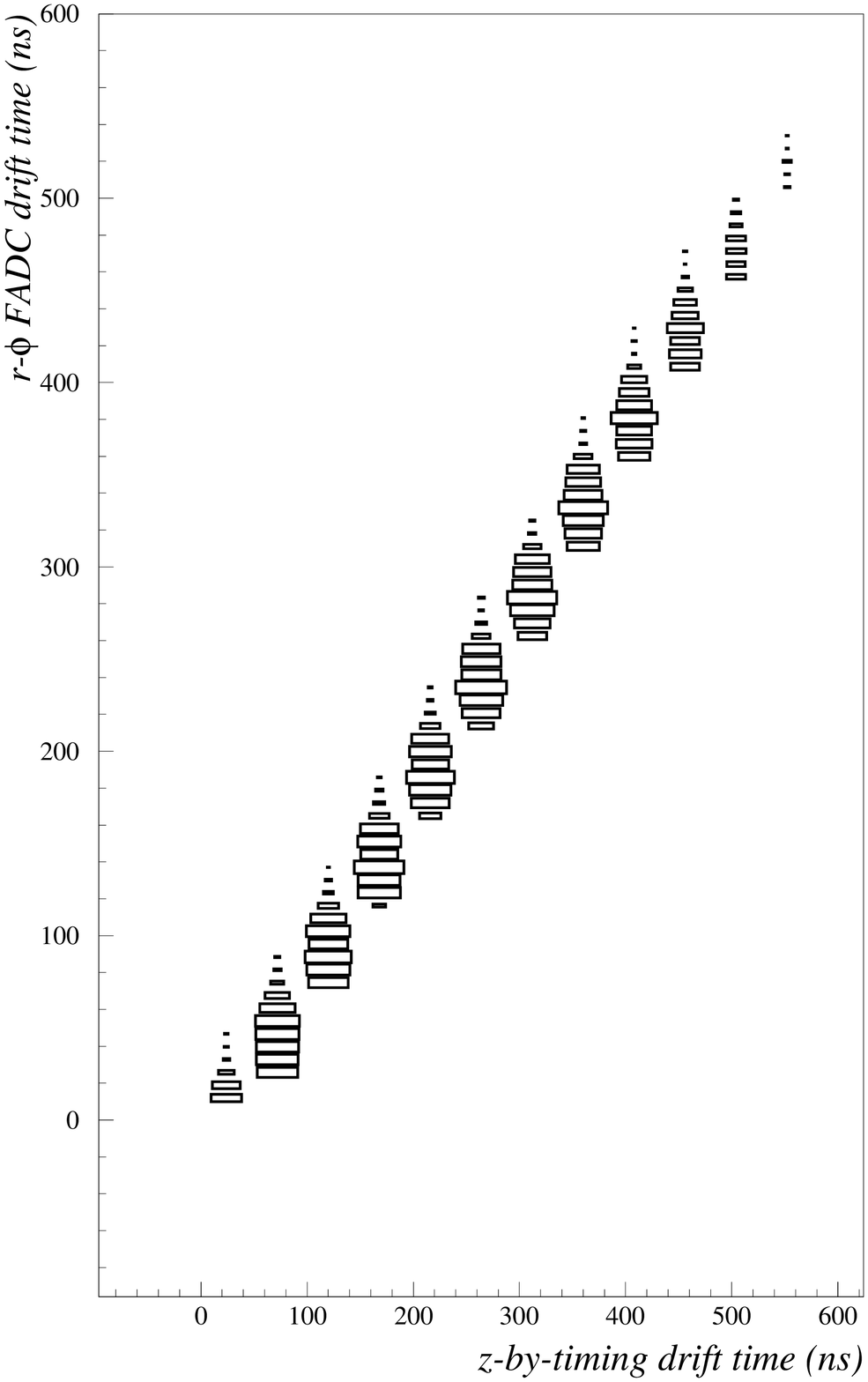,height=12cm}
\caption{
The correlation between $z$-by-timing and $r-\phi$ FADC drift times.}
\vfill
\label{fig:drift_FADC}
\end{center}
\end{figure}

Fig.~\ref{fig:drift_time} shows the
drift time of hits in bins of 48~ns measured by the $z$-by-timing system for
the eight wire-layers in Superlayer~1, taken from a sample of 
DIS events. The drift times have been corrected to correspond
to the same $t_0$.
All distributions have a sharp rise
at a drift time of zero (corresponding to a hit close to the sense
wire), and fall away at approximately 500~ns, the maximum drift time
within a cell. The extension to larger drift times as the wire-layer number
increases is a result of the tapered cell geometry. It should be
noted that outside the drift-time window 
the out-of-time background is very low. Fig.~\ref{fig:drift_FADC}
shows an example of the
correlation of $z$-by-timing and $r-\phi$ FADC drift times.
The background of mismatched
hits is negligible; less than 0.5\% of hits have a drift-time 
difference between the two systems
of greater than $\pm$48~ns (the $z$-by-timing digitization bin size). 
Events outside the drift-time window indicate the level
of random hits in the $z$-by-timing system is less than 1\%.

\subsection{Resolution and Efficiency}

The $z$-by-timing system is a vital component of the ZEUS
track trigger and hence it is important to understand its
efficiency and resolution, and to monitor its performance 
as a function of time.
These parameters have been studied utilizing the ZEUS track
reconstruction package VCTRAK~\cite{VCTRAK}.
The package uses full three dimensional stereo 
information provided by the $r-\phi$ FADC system and therefore, 
to first order, the reconstructed track fits 
are  independent of the hits
recorded by the $z$-by-timing system. 
Residuals and efficiencies of the $z$-by-timing hits
with respect to these tracks are presented below. 

\begin{figure}[htb]
\begin{center}
\epsfig{file=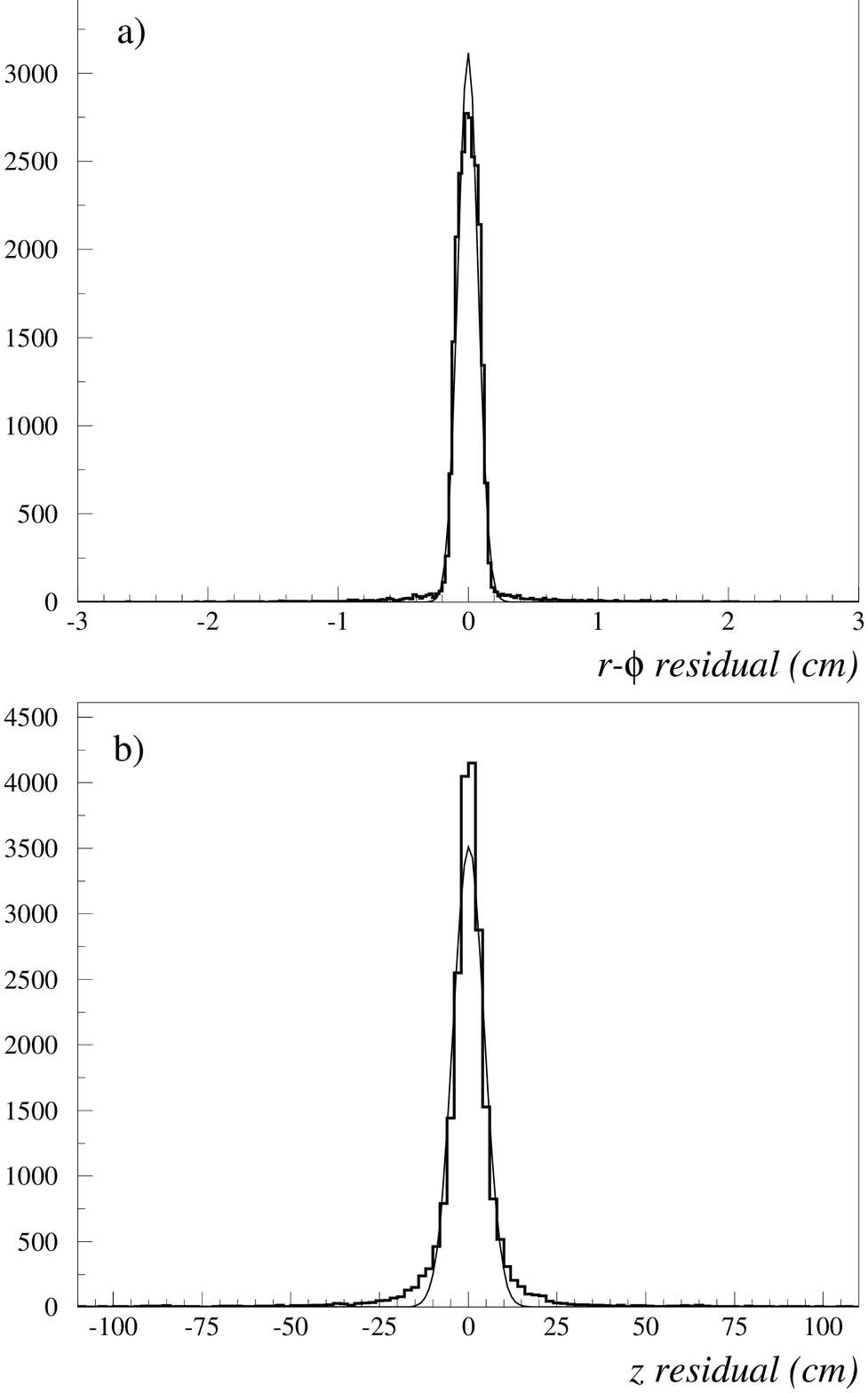,height=13cm}
\caption{
The resolution of the $z$-by-timing system~: the residuals (a) in
$r-\phi$ and (b) in $z$.}
\vfill
\label{fig:residuals} 
\end{center}
\end{figure}

The resolution of the $z$-by-timing system is demonstrated in 
Fig.~\ref{fig:residuals}. This shows residuals of the $z$-by-timing hits
from the fitted tracks in $r-\phi$ and $z$, integrated over 
tracks of all angles and momenta 
(where the $z$~measurements have been corrected offline
for the non-linear behaviour in the time-to-distance response using
the polynomial function of the form shown in Fig.~\ref{fig:Sshape}). 
Here it is important to explain how the 
resolutions and efficiencies are defined. Throughout this paper,
a hit is counted as efficient only if it exists, and has 
a measured $z$~value within $\pm$25~cm of its predicted value. This is
defined as being the limit beyond which a hit
is no longer useful. Similarly,
the $z$~resolution is measured from a Gaussian fit to the residual 
distribution performed
between the same limits, $\pm$25~cm, from the mean value.
Since the $z$~residual distribution is not perfectly Gaussian,
the fit can be biased by the tails of the distribution and hence is
sensitive to the limits over which it is performed. Therefore
the resolution and efficiency are not necessarily independent
parameters. 

As a result of the fits to Fig.~\ref{fig:residuals},
resolutions of 4.39$\pm$0.04~cm in~$z$ and 773$\pm$5~$\mu$m 
in $r-\phi$ are obtained.
The $z$~resolution of 4.4~cm is to be compared with
a previous test beam measurement of 3.0~cm
for single tracks, however which was made
in the absence of a magnetic  field and
using the preferred 50:50 argon-ethane gas mix~\cite{zbyt_3}.
A resolution of 4.4~cm corresponds to a time difference resolution
of approximately 350~ps.
The $r-\phi$ resolution of
780~$\mu$m compares with that expected from the quantization of
the drift-time measurement, namely 48~ns$\times v_d$/$\sqrt{\rm{12}}$
$\approx$700~$\mu$m, where the drift velocity ($v_d$) is 
taken to be 50~$\mu$m/ns.
The efficiency of the $z$-by-timing measurement, averaged over
tracks of all angles and momenta, is 0.90$\pm$0.01.

\begin{figure}[htb]
\begin{center}
\epsfig{file=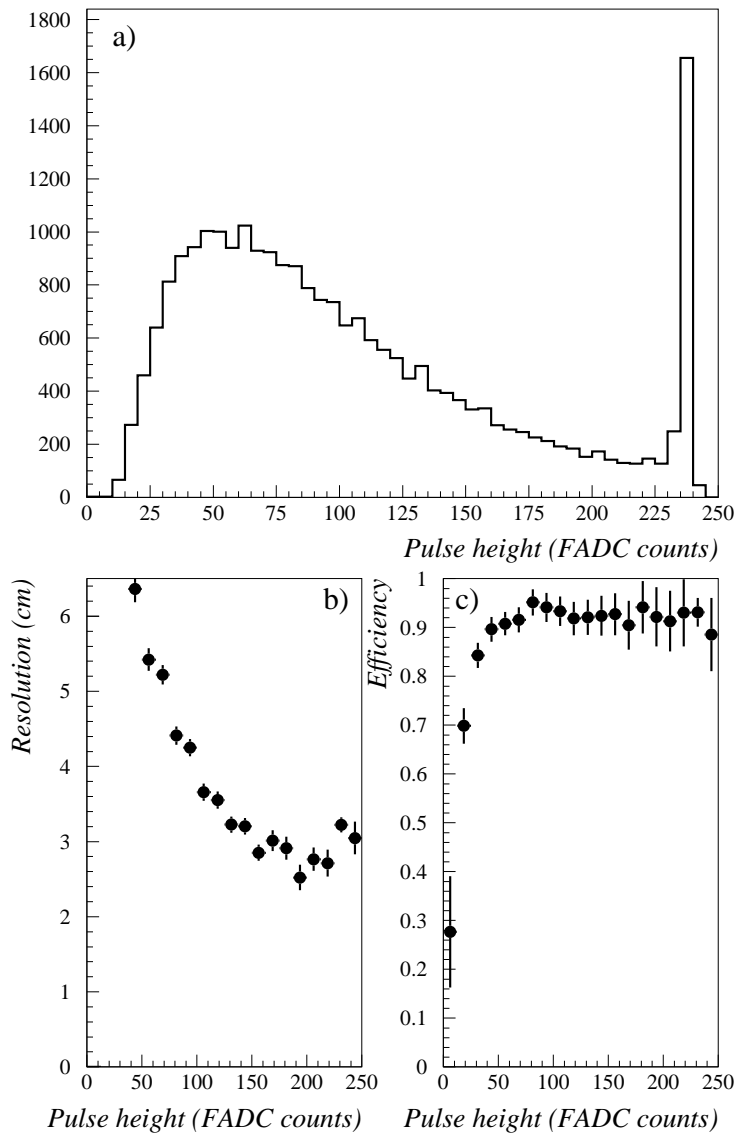,height=15cm}
\caption{
(a) The pulse height distribution, (b) the $z$~resolution and 
(c) the measurement efficiency as a function of pulse height.}
\vfill
\label{fig:res_ph} 
\end{center}
\end{figure}

It is well known that good signal to noise is 
essential for the $z$-by-timing technique~\cite{zbyt_2,zbyt_3}.
It is necessary to run the chamber at high gas gain in 
order to maximize
signal, however the need to minimize sense wire standing currents 
to preserve chamber longevity provides
the overriding constraint on its operation.
The requirement to enhance electron identification by
measuring ionization loss ($dE/dx$) in the CTD
also dictates the need to run at lower gain. Operation at a gas gain of
approximately 1$\times$10$^5$ has proved to be an acceptable
compromise, given normal background conditions of the HERA machine.
The uncorrected pulse height distribution in units of FADC counts
as measured by the $r-\phi$ FADC system at this gas gain is shown in 
Fig.~\ref{fig:res_ph}(a). Here the scale normalization is approximately
3~mV/count; the peak around 235 counts corresponds to pedestal-subtracted
FADC overflow.
Fig.~\ref{fig:res_ph}(b) confirms that the $z$~resolution 
depends strongly on pulse height, and hence on signal to noise. 
The resolution substantially improves as the mean pulse amplitude
increases,  and we achieve 3.0~cm in the limiting case
(where the residuals have again been calculated using tracks of all 
angles and momenta).
Fig.~\ref{fig:res_ph}(c) shows the $z$-by-timing hit efficiency 
as a function of pulse amplitude. 
Again as expected, the efficiency is correlated with pulse height.
For  pulses of small amplitude, the fall-off in efficiency is the result of
a 40~mV voltage threshold applied 
at the input to the $z$-by-timing readout cards
(corresponding to approximately 13~FADC counts). 

\begin{figure}[htb]
\begin{center}
\epsfig{file=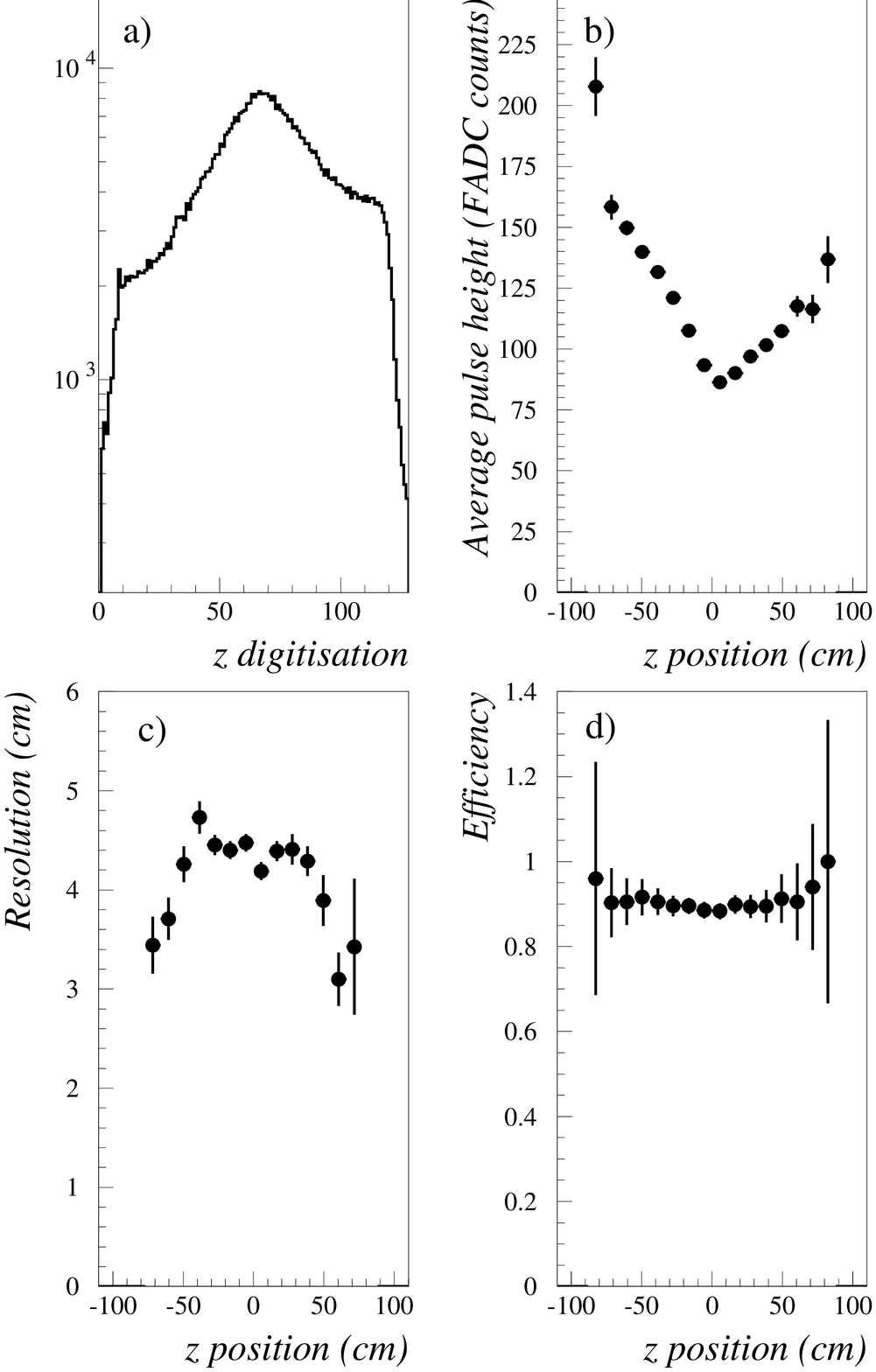,height=15cm}
\caption{
(a) The digitization (FADC counts) of $z$~position
of hits from the $z$-by-timing system for the DIS sample,
(b) the average $r-\phi$ FADC pulse height as a function
of  $z$~coordinate, (c) the $z$~resolution 
as a function of $z$~position along the wire,
and (d) the measurement efficiency.}
\vfill
\label{fig:z_digi} 
\end{center}
\end{figure}

Previous studies have shown that the $z$~resolution and
measurement efficiency are dependent on the $z$~coordinate of the hit on the 
wire~\cite{zbyt_2}. Since a track produced in an
$e-p$ collision nominally originates from the centre of the detector, its
path length through the chamber varies as
1/sin$\theta$, where $\theta$ is its polar angle.
This results in more primary ionization on average
arriving onto each sense wire,
hence greater pulse amplitude,  as the $|z|$~coordinate of the hit
increases away from the wire centre.
The distribution of 
uncorrected $z$~position, measured by the $z$-by-timing system
in units of FADC counts,
is shown in Fig.~\ref{fig:z_digi}(a) for the DIS data sample. 
Here the 2.03~m wire length is digitized between
0 (rear electron direction) and 127 (forward proton direction). The
detailed structure of the distribution depends on the nature of the
DIS cross section, the fragmentation process, and the cuts imposed.
It is interesting to note that the sensitive length of the chamber
starts about 6 counts in from the ends, which is due
to the calibration points being outside the active volume of the chamber
(see Section~\ref{sec:ZbyT}). 
The variation in the average pulse height as a function
of the  $z$~coordinate of the hit (measured by the $z$-by-timing system)
is demonstrated in Fig.~\ref{fig:z_digi}(b).
A marked asymmetry is observed in the distribution, 
caused by attenuation of pulses as they travel down
the length of the wire (pulse amplitudes are sampled only from the 
rear end of the CTD). Figs.~\ref{fig:z_digi}(c)
and (d) show the 
$z$~resolution and efficiency as a function of $z$~position.
As expected, the resolution improves at the wire ends due to improved
signal to noise of the larger pulses.
Note however that pulse reflections slightly degrade the resolution 
when a track passes close to the wire ends~\cite{zbyt_2}
and this can partially counteract the effect of improved signal.

\begin{figure}[htb]
\begin{center}
\epsfig{file=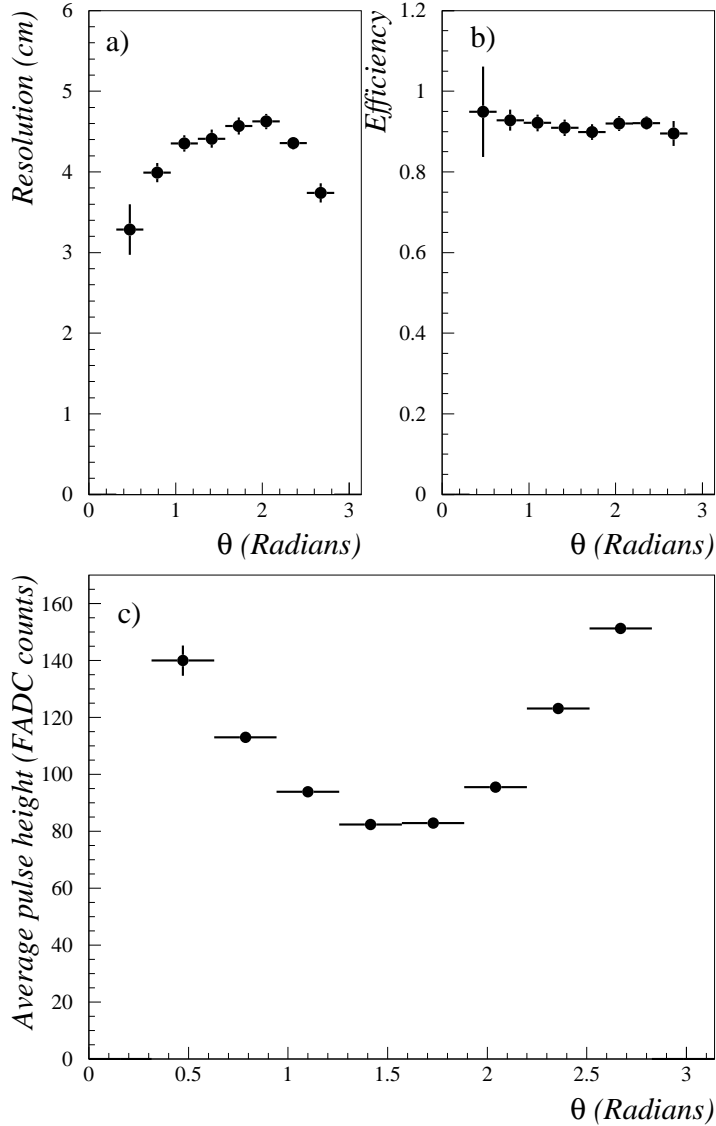,height=15cm}
\caption{
(a) The $z$~resolution, (b) the measurement efficiency
and (c) the mean pulse height as a function of polar angle, $\theta$.}
\vfill
\label{fig:res_theta} 
\end{center}
\end{figure}
\begin{figure}[htb]
\begin{center}
\epsfig{file=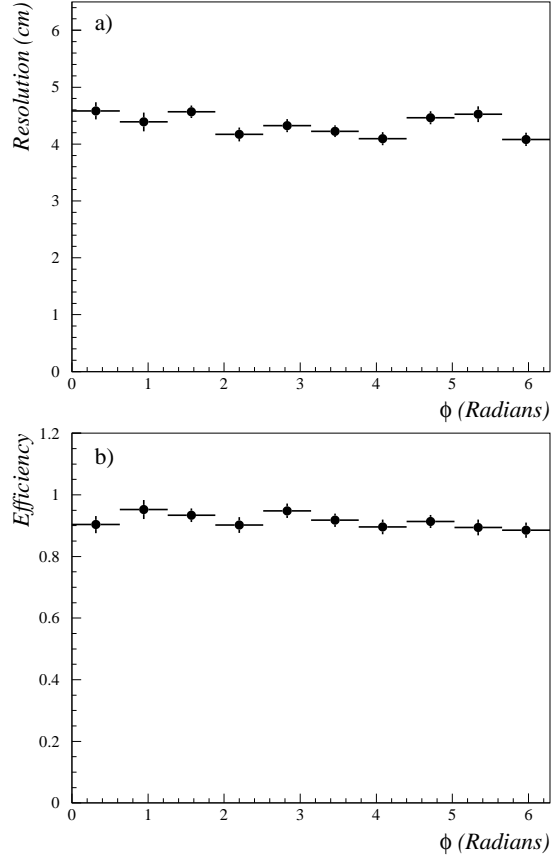,height=12cm}
\caption{
(a) The $z$~resolution and (b) the measurement 
efficiency as a function of track azimuthal angle,~$\phi$.}
\vfill
\label{fig:res_phi}
\end{center}
\end{figure}

The $z$~resolution and measurement efficiency 
are shown as a function of polar
angle $\theta$ in Fig.~\ref{fig:res_theta}. As expected, there is a
strong dependence of the resolution on this angle. 
This is in contrast to the dependence on track  azimuthal 
angle,~$\phi$; it can be seen from Fig.~\ref{fig:res_phi}
that the resolution and efficiency are $\phi$-independent.

\begin{figure}[htb]
\begin{center}
\epsfig{file=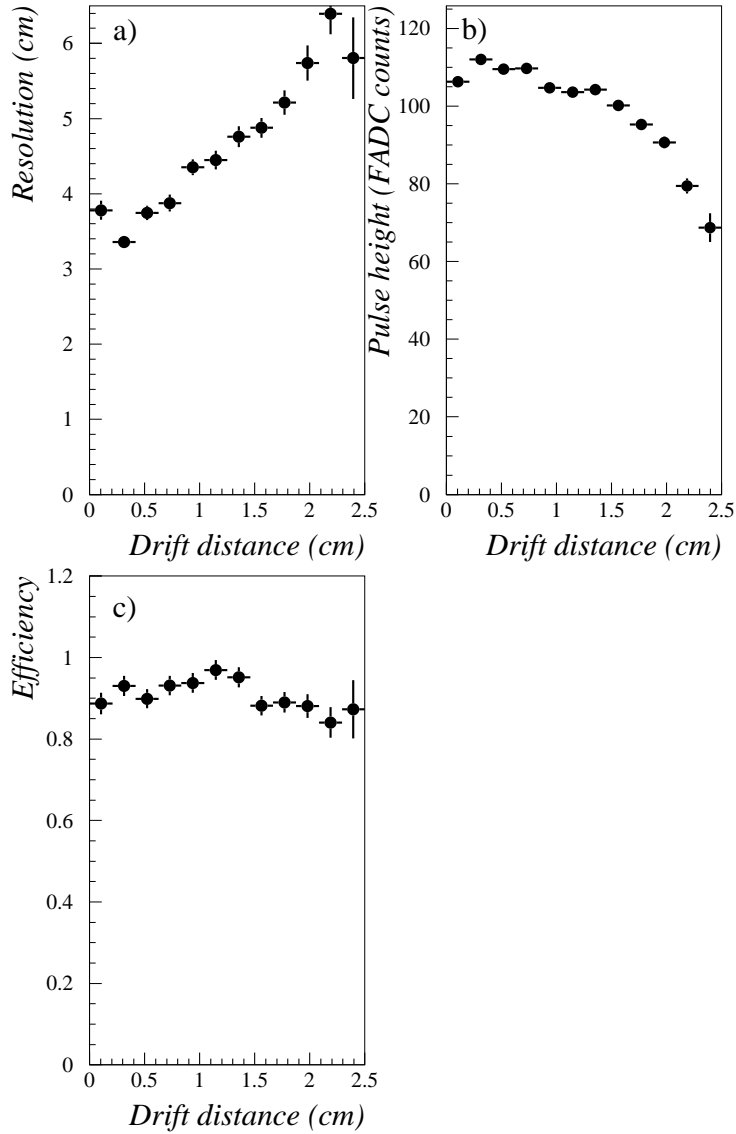,height=15cm}
\caption{
(a) The $z$~resolution, (b) the mean pulse height  and
(c) the measurement efficiency as a function of drift distance.}
\vfill
\label{fig:res_drift} 
\end{center}
\end{figure}

We have studied the dependence of $z$~resolution and measurement efficiency
as a function of the drift distance of hits from a sense wire.
Fig.~\ref{fig:res_drift}(a) shows the $z$~resolution, averaged over all
wires and track angles, as a function of
the drift time measured by the $r-\phi$ FADC system.
Ideally the dependence should be small, however it can be seen
that the resolution significantly degrades as the edges of the drift cells
are approached. This is understood in terms of two related effects. Firstly,
the CTD operated with a drift field close to
1.2~kV/cm during 1994 (see Section~\ref{sec:CTD}), 
and below this drift-field value field shaping becomes poor.
This results in a loss of prompt ionization reaching the sense wire,
especially at cell boundaries. 
Secondly, primary ionization is shared between neighbouring cells
close to a cell boundary and this also results in a loss of
signal at the sense wire.
The consequence of these effects is demonstrated in 
Fig.~\ref{fig:res_drift}(b)
which shows the mean pulse amplitude as a function of drift distance.
We observe that the degradation in resolution is indeed
correlated to a reduction of pulse height at the cell boundaries. 
The $z$-by-timing measurement efficiency as a function of drift distance
is shown in Fig.~\ref{fig:res_drift}(c).
Note that the effect of diffusion has a negligible effect
in such a small drift cell.

%

\begin{figure}[htb]
\begin{center}
\epsfig{file=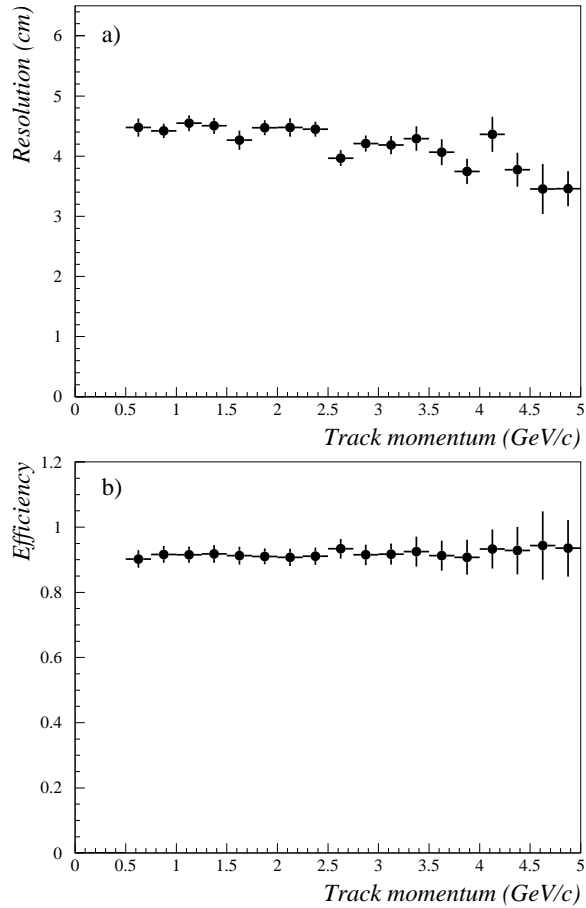,height=12cm}
\caption{
(a) The $z$~resolution and (b) the measurement 
efficiency as a function of pion track momentum.}
\vfill
\label{fig:res_p}
\end{center}
\end{figure}

The dependence of the $z$~resolution and measurement efficiency on track 
momentum is shown 
in Fig.~\ref{fig:res_p}. 
Since the tracks originate from quasi-elastic $\rho^0$ decay, both
tracks in the CTD are charged pions. It can be seen that there
is an improvement in resolution as pion momentum increases. This
is to be expected from the relativistic rise characteristic
of $dE/dx$ energy loss,
which results in increased  primary ionization and hence
improved performance.

\begin{figure}[htb]
\begin{center}
\epsfig{file=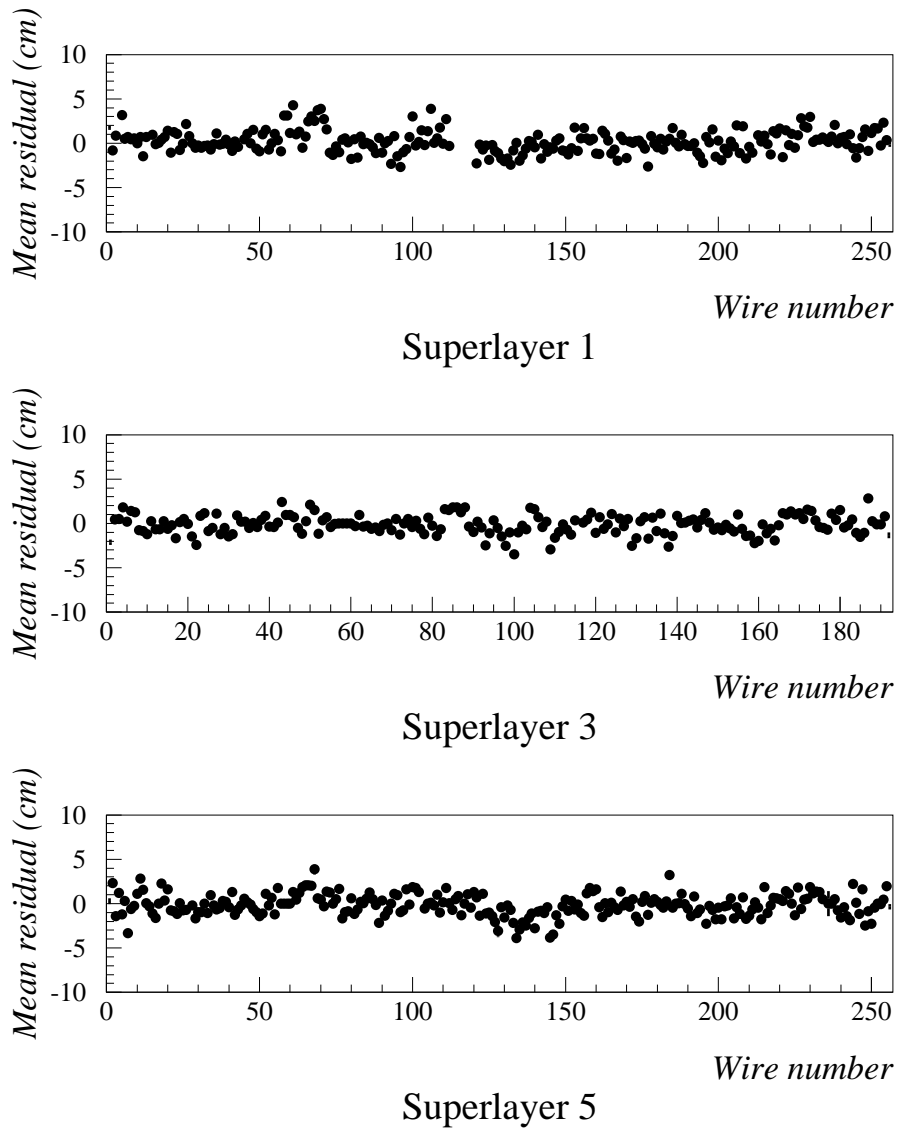,height=15cm}
\caption{
The mean of the residuals to the fitted tracks for each
wire as a function of wire number in each superlayer.}
\vfill
\label{fig:wiremean}
\end{center}
\end{figure}

A requirement in achieving good $z$-by-timing resolution is
maintaining an accurately calibrated system 
(i.e. a measurement of
0 and 127 FADC counts corresponding to hits at respective wire ends).
If either one of the end-points on any given wire is badly calibrated,
this will result in a shift of the mean of the residuals to the fitted tracks
for that wire.  
The mean residual values are
shown for each of the 704 wires in Fig.~\ref{fig:wiremean}
for the DIS data sample. It can be seen that the 
mean residuals generally have values close to zero, demonstrating that the 
calibration points are accurately known.
The RMS spread of these data about zero for all wires 
is close to 1~FADC count, and the same for each of Superlayers 1,~3 and 5.
This corresponds to an uncertainty in the position measurement
of approximately 1.5~cm, to be compared with
the measured $z$~coordinate resolution of 4.4~cm.
Although the effects of calibration point uncertainty are implicitly 
included in this quoted resolution, it can be seen that they
make rather a small contribution when subtracted in quadrature.

\begin{figure}[htb]
\begin{center}
\epsfig{file=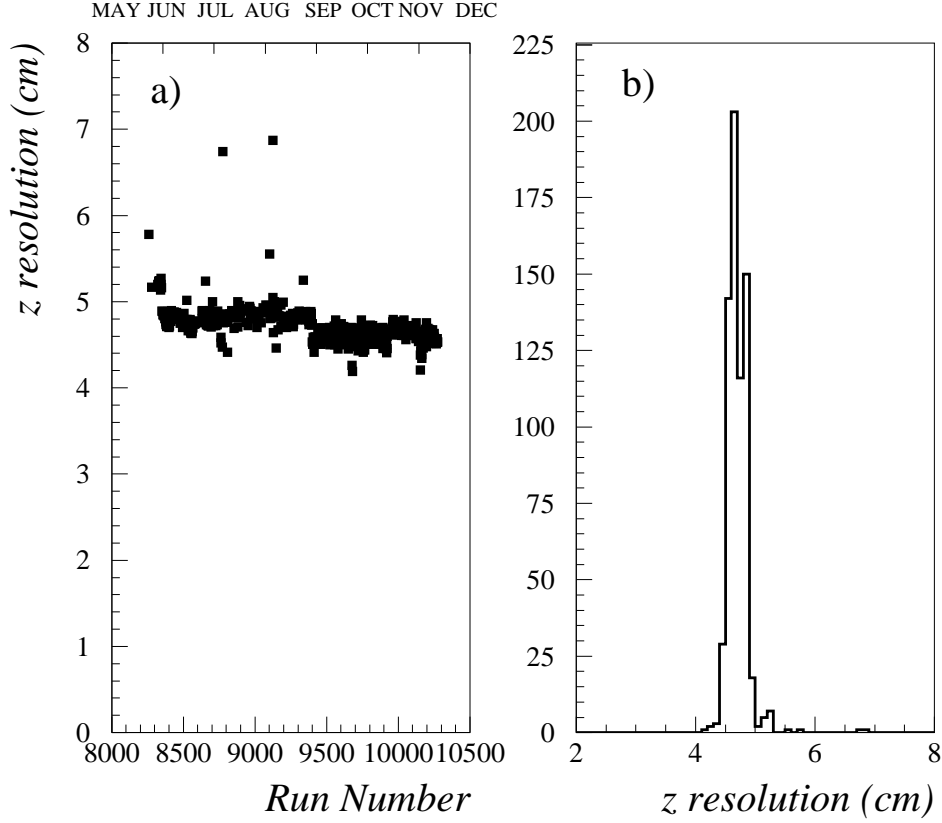,height=12cm}
\caption{
(a) The $z$~resolution as a function of ZEUS run number
and (b) its projection, output from the ZEUS TLT.}
\vfill
\label{fig:resrun} 
\end{center}
\end{figure}

It is essential to monitor
the performance of the $z$-by-timing system continuously over time.
In particular, variations in efficiency and resolution
would lead to instabilities in the ZEUS trigger (of which
the $z$-by-timing system is an integral component).
Monitoring is achieved during data-taking by 
the ZEUS Third Level Trigger which 
performs an (unoptimized) track fit from the $z$-by-timing data in real time.
Fig.~\ref{fig:resrun} shows the $z$~resolution from the TLT
as a function of run number during the 1994
data-taking period of ZEUS, which
lasted for approximately six months. It can be seen
that the $z$-by-timing system is stable over the
period; any systematic fluctuation can be understood in terms of small
changes  to the CO$_2$ concentration in the CTD gas mix.   

\subsection{Multihit Performance}

The high density of particles within jets in $e-p$ events at HERA 
necessitates good double-hit resolution of the front-end electronics.
As discussed in Section~\ref{sec:ZbyT}, the digitization time of the
$z$-by-timing system is 48~ns which, in principle,
defines the minimum time for sampling a second hit following the first.
 
A study has been made of the multihit performance of the 
$z$-by-timing system using the DIS data sample.
Again the procedure is to match measured $z$-by-timing hits
with those hits predicted by the tracking package.
Events are selected where two reconstructed tracks are identified 
as occupying the same cell (and hence two hits are expected on the same wire). 
The association of measured hits with a given track 
is made by matching drift times of the hits with the predictions. 
After hits have been matched, the efficiency 
(and the resolution) associated with the second hit can then be calculated. 

\begin{figure}[htb]
\begin{center}
\epsfig{file=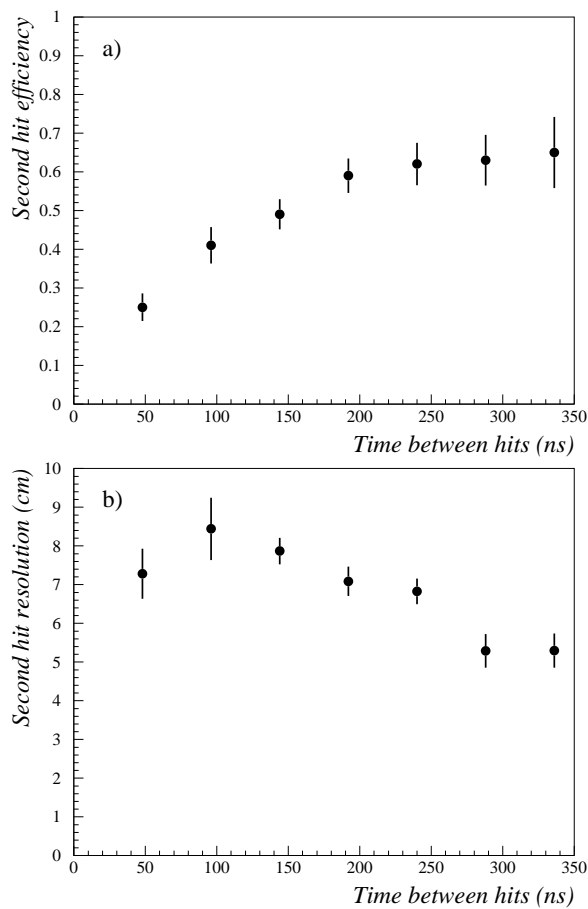,height=12cm}
\caption{
(a) The $z$-by-timing measurement efficiency and (b) the $z$~resolution
of the second hit as a function of its arrival time after the first.}
\vfill
\label{fig:Res_secondhit}
\end{center}
\end{figure}

The efficiency for observing the second hit is
shown as a function of the difference in the arrival times of the two hits
in Fig.~\ref{fig:Res_secondhit}(a). 
Here the time between the hits is directly calculated 
from the measured drift times in the case of two hits being observed,
and from the track predictions in the case of no second hit.
The efficiency of the second hit is relatively poor if it arrives
soon after the first, but gradually improves for later drift times.
A measurement of the resolution of the second hit is 
shown in Fig.~\ref{fig:Res_secondhit}(b). It can be seen that there is
a `recovery time' of $\sim$400~ns before the nominal resolution is 
restored. The resolution of the second hit
is degraded for small pulse separation
because the second pulse is often mis-shapen since
it sits on the trailing edge of the first. 

\section{Summary}
\label{sec:summary}

A $z$-by-timing readout system,
which provides fast three-dimensional space-point information,
has been developed for the ZEUS Central Tracking 
Detector. 
Its basis is a time difference
measurement using a Time-to-Amplitude Converter to 
determine the $z$~coordinate of hits.  
The HERA environment has necessitated
pipelined data storage with a fully customized method of timing control.
The $z$-by-timing system utilizes a calibration procedure which
has maintained stable, sub-nanosecond timing accuracy.
The information provided by the $z$-by-timing system
is used as input to all three levels of the ZEUS trigger
and provides independent, stand-alone 3-D hit readout.
For gas gains of approximately 1$\times$10$^5$,
we have achieved a $z$~coordinate resolution of 4.4~cm, averaged 
over  tracks of all angles and momenta.

\begin{figure}[htb]
\begin{center}
\epsfig{file=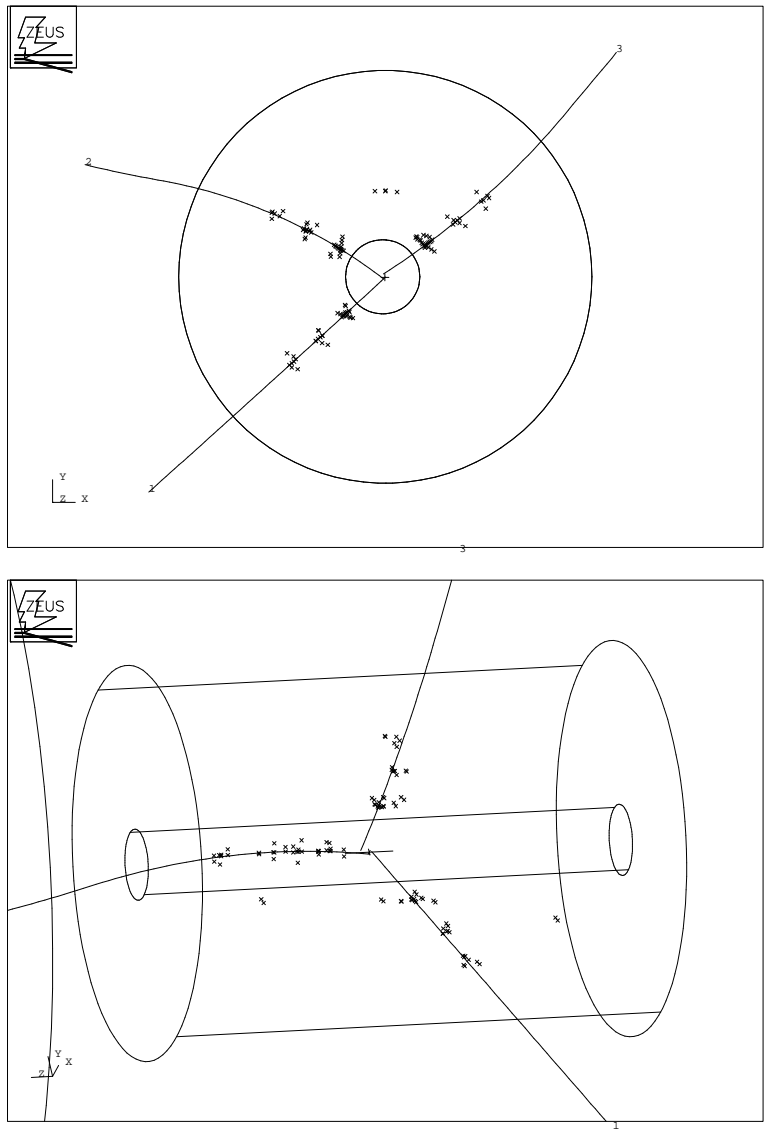,height=15cm}
\caption{
A DIS $e-p$ event recorded in the ZEUS CTD with the $z$-by-timing system.
A scattered electron and two decay muons from a diffractively 
produced J/$\psi$ are visible in the detector. 
Projections in $r-\phi$ and $r-z$ are shown, with offline reconstructed
tracks (from the $z$-by-timing hits alone) superimposed on the raw hits.
The left-right ambiguous (ghost) hits are also displayed.}
\vfill
\label{fig:event_pik}
\end{center}
\end{figure}

A display of a DIS event recorded in 
the ZEUS CTD is shown in Fig.~\ref{fig:event_pik}. Only
$z$-by-timing hits are shown.  Offline tracks,
reconstructed solely from the $z$-by-timing hits, are superimposed on the 
figure. The ability of the
system to provide full three-dimensional space-points is clearly demonstrated. 

\newpage

\begin{flushleft}
{\bf Acknowledgements} \\
\end{flushleft}

\noindent

We wish to thank the members of the ZEUS UK Collaboration who have all
helped in this work. 
We are particularly grateful to R.S.~Gilmore for his efforts on 
the calibration system
and to D.~Allen, S.~Berry, P.~Chorley, G.~Harris, D.A.~Hayes, 
P.~Morawitz, G.L~Salmon and P.~Roberts for their contributions towards
the hardware development. We also thank R.~Cranfield, G.J.~Crone,
G.~Hartner, Y.~Iga, K.~Long, N.A.~McCubbin, V.A.~Noyes, 
J.~Shulman and I.A.~Vine for invaluable online and offline software support. 
The work would not have been possible without the 
achievement of the HERA machine group in providing
$e-p$ beams. Finally, we gratefully
acknowledge the financial support provided by
the UK Particle Physics and Astronomy Research Council.

\end{document}